\documentclass[preprint,aps,pdftex,nofootinbib,amsmath,amssymb,
amsfonts,superscriptaddress]{revtex4}
\usepackage{graphicx}
\usepackage{amsmath, amssymb, bm}   % AMS packages
\usepackage[dvips]{color}
\usepackage{epsfig}
\usepackage{natbib}
\usepackage{hyperref}
\usepackage{slashed}
\usepackage{epsfig}
\usepackage{color}
\usepackage{cancel}
\usepackage{ulem}
\newcommand{\simle}{\hspace*{0.2em}\raisebox{0.5ex}{$<$}
     \hspace{-0.8em}\raisebox{-0.3em}{$\sim$}\hspace*{0.2em}}

\begin{document}

\preprint{CTP-SCU/2017008}

\title{The Two-Nucleon $\bm{^1S_0}$ Amplitude Zero\\
in Chiral Effective Field Theory}

\author{M. S\'anchez S\'anchez}
\affiliation{Institut de Physique Nucl\'eaire, CNRS/IN2P3, \\
Univ.~Paris-Sud, Universit\'e Paris-Saclay, 91406 Orsay, France}
\author{C.-J. Yang}
\affiliation{Institut de Physique Nucl\'eaire, CNRS/IN2P3, \\
Univ.~Paris-Sud, Universit\'e Paris-Saclay, 91406 Orsay, France}
\author{Bingwei Long}
\affiliation{Center for Theoretical Physics, Department of Physics, 
Sichuan University, 29 Wang-Jiang Road, Chengdu, Sichuan 610064, China}
\author{U. van Kolck}
\affiliation{Institut de Physique Nucl\'eaire, CNRS/IN2P3, \\
Univ.~Paris-Sud, Universit\'e Paris-Saclay, 91406 Orsay, France}
\affiliation{Department of Physics, University of Arizona, Tucson, AZ 85721, 
USA}
\date{\today}

\begin{abstract}
We present a new rearrangement of short-range interactions
in the $^1S_0$ nucleon-nucleon channel within Chiral Effective Field Theory.
This is intended to reproduce the amplitude zero (scattering momentum
$\simeq 340~\mathrm{MeV}$) at leading order, and
it includes subleading corrections perturbatively in
a way that is consistent with renormalization-group invariance.
Systematic improvement is shown at next-to-leading order,
and we obtain results that fit empirical phase shifts remarkably well
all the way up to the pion-production threshold.
An  approach in which pions have been integrated out is included, which allows
us to derive analytic results that also fit phenomenology surprisingly well.
\end{abstract}

\date{\today}
\maketitle

%===================================================
\section{introduction}
\label{sec:intro}
%===================================================

The nuclear effective field theory (EFT) program
\cite{VanKolckBedaque02,ModTheNucFor}
conceives nuclear physics as the renormalization-group (RG)
evolution of Quantum Chromodynamics (QCD) at low energies,
formulated in terms of effective degrees
of freedom (nucleons, pions, \textit{etc.}).
The link with QCD written in terms
of more fundamental objects (quarks and gluons)
is ensured by imposing QCD symmetries
(particularly approximate chiral symmetry) as the only constraints on the
otherwise most general EFT Lagrangian.
Power counting (PC) rules tell which terms in this
Lagrangian (out of an infinite number)
should be taken into account when computing observables at a given order
in an expansion in powers of the small parameter $Q/M_{\rm hi}$,
where $Q$ is the characteristic external momentum of a process
and $M_{\rm hi} \simle M_{\rm QCD}\sim 1$ GeV is the EFT breakdown scale.
Thanks to the recent development of \textit{ab initio} methods,
which bridge the gap between
nuclear forces and currents on one hand and
nuclear structure and reactions on the other,
Chiral EFT ($\chi$EFT)
\cite{VanKolckBedaque02,ModTheNucFor,Machleidt'n'Entem}
is now better exploited than ever.
However, problems remain in the formulation of this EFT,
some of which we address here in the simplest, yet surprisingly
challenging, two-nucleon ($N\!N$) channel ---
the spin-singlet, isospin-triplet $S$ wave, $^1S_0$.

The initial applications of $\chi$EFT followed a scheme
suggested by Weinberg \cite{Weinberg90, Weinberg91} and Rho \cite{Rho:1990cf},
where a PC dictated by
naive dimensional analysis (NDA) \cite{Manohar:1983md,Georgi93} was assumed
to apply to the nuclear potential and currents.
The truncated potential is inserted into a dynamical equation
--- Lippmann-Schwinger (LS), Schr\"odinger, or one of their variants
for the many-body system ---
from whose exact solution nuclear wave functions are obtained.
Averages of
the appropriate, truncated currents give rise to scattering
amplitudes when the system is probed by external particles such as
photons or pions.
To deal with the singular nature of the potential and currents, an arbitrary
regularization procedure must be introduced. Unfortunately, already at
leading order (LO) NDA does not yield all the short-range interactions
necessary for the $N\!N$ amplitude to be approximately independent of the
regulator choice \cite{KSW96,NTvK,PavonValderrama:2005uj}.
Similar issues appear at higher orders \cite{YangElsterPhillips,Ya09B,ZE12} 
and also affect electromagnetic currents  \cite{Valderrama:2014vra}.
Given that non-perturbative renormalization can differ significantly from
the perturbative renormalization used to infer NDA, it is perhaps unsurprising
that a scheme based solely on NDA fails to produce
nuclear amplitudes consistent with RG invariance.

This problem appears even in $N\!N$ scattering in the
$^1S_0$ channel, where one-pion exchange (OPE) has a delta-function singularity
in coordinate space.
While NDA prescribes that the contact term, which supplements OPE in the LO
potential, is chiral-invariant, renormalization demands that a
chiral-symmetry-breaking short-range interaction also be present \cite{KSW96}.
According to NDA, such a chiral-breaking interaction,
being proportional to two powers of the pion mass,
should not appear
before two more orders (next-to-next-to-leading order, or N$^2$LO)
in the $Q/M_{\rm hi}$ expansion.
This ``chiral inconsistency'' motivated Kaplan, Savage, and
Wise \cite{KSW98bis, KSW98} to propose a PC where pion exchanges are treated
as perturbative corrections starting at next-to-leading order (NLO).
However, higher-order calculations soon made clear that such an approach
is not valid at low momenta in certain partial waves \cite{FMS00}.
The alternative is to treat OPE as LO only in the lower waves
\cite{NTvK,Birse:2005um,Valderrama:2009ei,Valderrama:2011mv,Long:2011qx,
Long:2011xw,Long:2012ve,Song:2016ale},
where suppression by the centrifugal barrier is not effective.
The angular-momentum suppression factor has been
studied recently in peripheral spin-singlet channels \cite{PeripheralSinglets}.

The $^1S_0$ partial wave was excluded from the analysis in
Ref. \cite{PeripheralSinglets}
because this particular channel presents,
in addition to the above renormalization issue, other features
that are not completely understood.
The situation has not improved greatly since the late 90s,
despite considerable effort
\cite{Kaplan97,Cohen:1998,Steele:1998zc,Mehen:1999,Frederico:1999,
Gegelia:1999,Kaplan:1999qa,Hyun:2000,Lutz00,Beane:2001bc,Nieves:2003,Oller:2003,
ValArr:2004,ValArr:2004bis,Frederico:2005,PA06,YangHuang,EntArrValMach,
SotoTarrus08,Shukla:2008sp,YangPhillips,Ya09B,Birse:2010jr,
Harada:2011,Long:2012ve,AndoHyun12,Szpigel:2012,Long13,Harada:2013,
Epelbaum-2015sha,Ren-2016jna}. 
Some of this work has been reviewed recently in Refs. \cite{Long16,Valderrama}.

A unique feature of this channel, which was recognized early on,
is fine tuning in the form of a very shallow virtual bound state.
OPE is characterized by two scales,
its inverse range given by the pion mass $m_\pi$ and
its inverse strength given by
$M_{N\!N} \equiv 16\pi f_\pi^2/(g_A^2 m_N) = \mathcal{O}(f_\pi)$,
where $m_N = \mathcal{O}(M_{\rm QCD})$ is the nucleon mass,
$f_\pi= \mathcal{O}(M_{\rm QCD}/(4\pi))$ is the pion decay constant,
and $g_A = \mathcal{O}(1)$ is the axial-vector coupling constant.
At the physical pion mass $m_\pi\approx 140$ MeV,
the virtual state's binding momentum $\aleph \sim 10$ MeV is much smaller
than the pion scales, and can only be
reproduced at LO through a fine tuning of the short-range interaction.
Physics of this state can be described simply
by another successful, renormalizable EFT, Pionless
(or Contact) EFT ($\slashed{\pi}$EFT).
In the very-low-energy regime of nuclear physics,
$Q\ll m_\pi$, pion exchange cannot be resolved, the EFT Lagrangian
contains only contact interactions, and the
two-body amplitude reduces
\cite{vanKolck:1997ut,KSW98bis,KSW98,VanKolck98}
to the effective range expansion (ERE).
To simultaneously capture physics at $Q\sim m_\pi$, however,
pion exchange
%\lbwcmt{OPE} 
%We are not talking LO only here, so all pion exchange needs to be retained
needs to be retained.
The perturbative expansion in $Q/M_{N\!N}$ prescribed by
Refs. \cite{KSW98bis,KSW98} converges very slowly, if at all,
in the low-energy region \cite{Beane:2001bc},
which leads to the identification of $M_{N\!N}$ as a low-energy scale
$M_{\mathrm{lo}}$, just as suggested by NDA.

Yet, it is disturbing that the NDA-prescribed LO potential produces
$^1S_0$ phase shifts that show large discrepancies with the Nijmegen
partial-wave
analysis (PWA) \cite{PWA,nnonline} even at moderate scattering energies.
In Ref. \cite{Long:2012ve} it was shown that ---
again at variance with NDA --- the first correction
in this channel appears already at NLO, in the form of a
contact interaction with two derivatives.
Still, only about half of the energy dependence of the amplitude
near threshold is accounted for at LO, so Ref. \cite{Long13} went a step further
by suggesting the promotion to LO of an energy-dependent
short-range interaction that reproduces the effective range
--- a generalization of the same suggestion for $\slashed{\pi}$EFT
\cite{BeaneSavage}.
Even this promotion leaves
significant room for improvement when compared to the Nijmegen PWA.
In particular, the empirical $^1S_0$ phase shift, thus the  amplitude,
vanishes at a center-of-mass
momentum $k = k_0\simeq 340~\mathrm{MeV}$.
Since $k_0$ is significantly below the expected breakdown scale
$M_{\rm QCD}$, we should consider it as
a soft scale where the EFT converges.
In contrast, we find that the LO phase shift of Ref. \cite{Long13} is around
$25^\circ$ at $k=k_0$ and does not vanish until $k$ reaches a few GeV.
Since higher orders need to overcome LO, 
convergence at momenta $k\sim k_0$ will be at best very slow.
This can only be remedied if LO contains the amplitude zero.
As pointed out in Ref. \cite{VanKolck98}, a low-energy zero requires
a different kind of fine tuning than the one that gives rise to
a shallow bound state.
When the zero appears at very low energies,
a contact EFT can be devised
(the ``other unnatural EFT'' of Ref. \cite{VanKolck98})
which gives rise to a perturbative expansion of the amplitude.
Such an expansion around $k=k_0$ in the presence of pions
was developed in Ref. \cite{Lutz00}.

Here we propose a rearrangement of the short-range part of $\chi$EFT that leads
to the existence of the amplitude zero at LO, in addition to the shallow
virtual state.
The PC of Ref. \cite{VanKolck98} is generalized with the purpose of
including the non-perturbative region that contains the virtual state.
This is patterned on an idea
originally developed for doublet neutron-deuteron ($nd$) scattering at very
low energies \cite{nd}, where the amplitude has a zero at small imaginary
momentum, in addition to a shallow virtual state.
We develop an expansion in $Q/M_{\rm hi}$ for $Q\sim M_{\rm lo}$,
which gives a renormalizable amplitude order by order.
Following a successful approach to $\slashed{\pi}$EFT \cite{Konig:2015aka},
the virtual state is assumed to be located right at threshold
at LO and is moved to a binding momentum
$\sim  M_{\mathrm{lo}}^2/M_{\mathrm{hi}}$ at NLO.
We calculate NLO corrections and show a systematic improvement
in the description of the phase shift.

A challenging feature of $\chi$EFT is that it usually
does not yield analytical expressions for amplitudes.
In order to facilitate an understanding of the properties of
the $N\!N$ amplitude, we also consider a
version of our PC for the theory without
explicit pions, where we retain $k_0\sim M_{\rm lo}$ but take $M_{N\!N}\to \infty$.
To our surprise, even though $k_0 > m_\pi$, this new version of
$\slashed{\pi}$EFT also produces a good description of
the empirical phase shifts.

Our approach is in line with Refs. \cite{Kaplan:1999qa,Birse:2010jr},
which argued that short-range forces in the spin-singlet $S$ wave
must produce rapid energy
dependence.
It is a systematic extension of the potential proposed in Ref. \cite{Kaplan97},
and it resembles the unitarized approach of Ref. \cite{Lutz00}.
More generally, it can be seen as the EFT realization of
Castillejo-Dalitz-Dyson (CDD) poles \cite{CDD} in $S$-matrix theory.
Traditional $S$-matrix tools, such as the $N/D$ method, have
recently received renewed attention in the $N\!N$ system
(\textit{e.g.} Ref. \cite{OllerEntem}).
The $D$ function is determined modulo the addition of CDD poles, which
result in zeros of the scattering amplitude.
In particular, the momentum $k_0$ may be identified with the position of
a CDD pole in the $^1S_0$ channel \cite{Krivoruchenko}.
An EFT provides a systematic description of the two-body CDD pole
that can be naturally extended to more-body systems.

This article is structured as follows.
In Sec. \ref{sec:pionless}
we present an initial approach (``warm-up'') to the problem
on the basis of a modified organization of $\slashed{\pi}$EFT up to NLO.
The proposed PC is discussed in detail, and
RG invariance is demonstrated explicitly.
In Sec. \ref{sec:pionful}  we bring OPE into LO; also, we compare with the 
results \cite{nnonline} of the high-quality
\textit{Nijm93} potential \cite{Nijm93} before and after the inclusion of 
the NLO potential in this $\chi$EFT.
Conclusions and outlook are presented in Sec. \ref{sec:conclusion}.

%============================================================
\section{Pionless Theory}
\label{sec:pionless}
%============================================================

Our first approach to the problem will omit explicit pion exchange
(and also electromagnetic interactions, which are small
for $k\gtrsim10~\mathrm{MeV}$ anyway,
as well as other small isospin-breaking effects \cite{Konig:2015aka}).
This will allow us to find analytical results, which cannot be reached if
one includes OPE in (fully iterated) LO.

In the absence of explicit pions and nucleon excitations,
all interactions among nucleons are of contact type.
The part of the ``standard'' $\slashed{\pi}$EFT Lagrangian relevant for
the $N\!N$ $^1S_0$ channel is
\begin{equation}
\mathcal{L}_{\slashed{\pi}}^{(\rm ct)} = N^\dagger\left(i\partial_0
+ \frac{\bm{\nabla}^2}{2m_N}\right)N
- C_0\left(N^T \vec{P}_{^1S_0} N\right)^\dagger\cdot
\left(N^T \vec{P}_{^1S_0} N\right) + \cdots
\label{Lpiless},
\end{equation}
where $N$ is the isodoublet, bispinor nucleon field
and the $N\!N$ $^1S_0$ projector is expressed in terms of the Pauli matrices
$\bm{\sigma}$ $(\vec{\tau})$ acting on spin (isospin) space as
$\vec{P}_{^1S_0} = \sigma_2 \vec{\tau} \tau_2/\sqrt{8}$,
while ``$\cdots$'' means more complicated interactions
and relativistic corrections suppressed
by negative powers of the breakdown scale of the theory.
Now, the interaction term in Eq. \eqref{Lpiless} may be rewritten if,
following Ref. \cite{Kaplan97}, an auxiliary ``dibaryon'' field $\vec{\phi}$
with quantum numbers of an isovector pair of nucleons is introduced,
\begin{align}
- \,C_0 \left(N^T \vec{P}_{^1S_0} N\right)^\dagger\cdot
\left(N^T \vec{P}_{^1S_0} N\right) \,
\leftrightarrow \,
\vec{\phi}^{\,\dagger} \cdot \Delta\vec{\phi}
- g \left(\vec{\phi}^{\,\dagger} \cdot N^T \vec{P}_{^1S_0}N
+ \mathrm{H.c.}\right).
\label{onedibaryon}
\end{align}
The dibaryon residual mass $\Delta$ and the dibaryon-$N\!N$ coupling
$g$  are such that $C_0 = g^2/\Delta$,
as can be straightforwardly checked if one performs the corresponding Gaussian
path integral. This parameter redundancy permits the convenient
choice \cite{GSS08}
\begin{equation}
g^2 \equiv \frac{4\pi}{m_N}.
\label{conv}
\end{equation}
Higher-order contact interactions
can be reproduced by the inclusion
of the dibaryon's kinetic term and derivative dibaryon-$N\!N$ couplings.

The standard PC of $\slashed{\pi}$EFT
\cite{vanKolck:1997ut,KSW98bis,KSW98,VanKolck98}
accounts for the presence of a shallow virtual state at LO,
but does not produce as much energy dependence as the phenomenological
phase shifts.
A promotion of the dibaryon kinetic term to LO \cite{BeaneSavage}
allows for the reproduction of the derivative of the amplitude
with respect to the energy around threshold, but it is unable to
generate the amplitude zero
by itself. This is not a problem in the context of standard
$\slashed{\pi}$EFT, since $k_0$ --- numerically larger than $m_\pi$ --- is
presumably outside the scope of this theory.
But here we aim at reformulating the theory in a way such that $k_0$
is considered below the breakdown scale, so as to illustrate the proposed
reformulation of the $\chi$EFT PC in Sec. \ref{sec:pionful}.

Inspired by an EFT for $nd$ scattering at very low energies
\cite{nd}, we consider here a generalization with {\it two} such
dibaryon fields, $\vec{\phi}_{1,2}$,
\begin{eqnarray}
\mathcal{L}_{\slashed{\pi}}^{(2\phi)} &=&
N^\dagger\left(i\partial_0 + \frac{\bm{\nabla}^2}{2m_N}\right)N
+ \sum_{j=1,2}{\vec{\phi}}_j^{\,\dagger}
\cdot\left[\Delta_j+c_j\left(i\partial_0+\frac{\bm{\nabla}^2}{4m_N}\right)\right]
\vec{\phi}_j
\notag\\
&&-\, \sqrt{\frac{4\pi}{m_N}}\sum_{j=1,2}
\left({\vec{\phi}}_j^{\,\dagger} \cdot N^T \vec{P}_{^1S_0}N + \mathrm{H.c.}\right)
+\cdots,
\label{L}
\end{eqnarray}
where we have made use of Eq. \eqref{conv}
and displayed explicitly the kinetic dibaryon terms with
dimensionless factors $c_j$.
As we will see, such an extension naturally allows us to reproduce
the amplitude zero already at LO, greatly improving the description
of the empirical phase shifts.

\begin{figure}[tb]
\centering
\includegraphics[scale=.2]{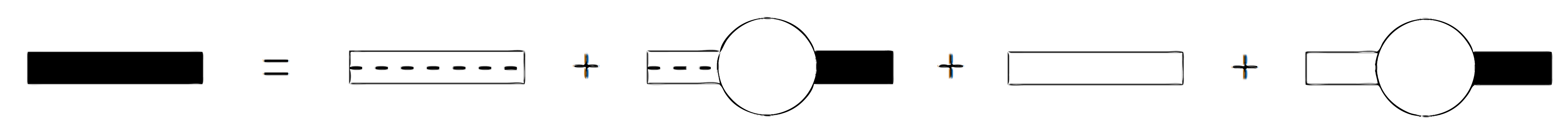}
\caption{Full two-dibaryon propagator (solid box) resulting from
the non-perturbative dressing of bare dibaryon-1
(dashed box) and dibaryon-2 (plain box) propagators
with nucleon bubbles (circles).}
\label{dibaryonprop}
\end{figure}

To illustrate the effects of the two dibaryons, we
neglect for now the interactions represented by ``$\cdots$'' in Eq. \eqref{L}.
At momentum $k=\sqrt{m_NE}$,
where $E$ is the center-of-mass energy,
the on-shell $T$ matrix is written
in terms of the $S$ matrix and the phase shift $\delta$ as
\begin{equation}
T(k) = \frac{2\pi i}{m_N k} \left[S(k) -1 \right]
=\frac{4\pi}{m_N}\left[-k\cot\delta(k)+ik\right]^{-1}.
\label{phaseshift}
\end{equation}
Loops are regularized by a
momentum cutoff $\Lambda$ in the range $\Lambda\gtrsim M_{\mathrm{hi}}\gg k$
and a regulator function
$f_{\mathrm{R}}(q/\Lambda)$, with $q$ the magnitude of the off-shell nucleon
momentum, that satisfies
\begin{equation}
f_{\mathrm{R}}(0) = 1, \quad
f_{\mathrm{R}}(\infty) = 0. \label{regulator}
\end{equation}
Computing the two-dibaryon self-energy,
\textit{i.e.} dressing up the bare two-dibaryon propagator
\begin{equation}
\mathcal{B}(k;\Lambda) =
\sum_{j}\left(\Delta_j(\Lambda)+c_j(\Lambda)\,\frac{k^2}{m_N}\right)^{-1}
\equiv \frac{m_N}{4\pi}\,V(k;\Lambda)
\label{B}
\end{equation}
with nucleon loops
(see Fig. \ref{dibaryonprop}), yields
\begin{equation}
\mathcal{D}(k;\Lambda) =
\left(\frac{1}{\mathcal{B}(k;\Lambda)}+\mathcal{I}_0(k;\Lambda)\right)^{-1}
\equiv \frac{m_N}{4\pi}\,T(k;\Lambda) .
\label{D}
\end{equation}
In this equation we introduced the regularized integral
\begin{equation}
\mathcal{I}_0(k;\Lambda) =
4\pi\int\frac{\mathrm{d^3}q}{\left(2\pi\right)^3}
\, \frac{f_{\rm{R}}(q/\Lambda)}{q^2-k^2-i\epsilon}
= ik + \theta_{1}\Lambda
+ \frac{k^2}{\Lambda}\sum_{n=0}^\infty \theta_{-1-2n}
\left(\frac{k}{\Lambda}\right)^{2n},
%\sum_{n=0}^\infty \theta_{1-2n}\frac{k^{2n}}{\Lambda^{2n-1}},
\label{I0}
\end{equation}
where the dimensionless coefficients $\theta_{n}$ depend on the specific
regularization employed.
For example, for a sharp-cutoff prescription with a step function
it turns out that $\theta_{n}=2/(n\pi)$,
while in dimensional regularization with minimal subtraction
we have simply $\theta_{n}=0$.
We thus arrive at
\begin{eqnarray}
\left[\frac{m_N}{4\pi}\, T(k;\Lambda)\right]^{-1} &=&
\frac{\left[\Delta_1(\Lambda)+c_1(\Lambda)\, k^2/m_N\right]
\left[\Delta_2(\Lambda)+c_2(\Lambda)\, k^2/m_N\right]}
{\Delta_1(\Lambda)+\Delta_2(\Lambda)
+\left[c_1(\Lambda)+c_2(\Lambda)\right]\, k^2/m_N}+ik
\nonumber\\
&&+\,\theta_1\Lambda+\theta_{-1}\frac{k^2}{\Lambda}
+\mathcal{O}\left(\frac{k^4}{\Lambda^3}\right).
\label{4param}
\end{eqnarray}

When $k$ is much smaller than any other scale,
this inverse amplitude
reduces at large cutoff to the ERE,
\begin{equation}
\left[\frac{m_N}{4\pi}T(k)\right]^{-1}
= \frac{1}{a} + ik - \frac{r_0}{2} k^2 - \frac{P_0}{4}k^4
+\cdots,
\label{ERE}
\end{equation}
where, for neutron-proton ($np$) scattering,
$a\simeq -23.7\,\mathrm{fm} \simeq -(8\,\mathrm{MeV})^{-1}$
\cite{KoesterNistler75}
is the scattering length,
$r_0\simeq 2.7\,\mathrm{fm} \simeq  (73\,\mathrm{MeV})^{-1}$ \cite{Lomon1974}
is the effective range,
$P_0\simeq 2.0\,\mathrm{fm^3} \simeq  (158\,\mathrm{MeV})^{-3}$ \cite{PB09}
is the shape parameter,
and so on.
In addition, Eq. \eqref{4param} allows for a pole at a momentum
$k_0\simeq 340~\!\mathrm{MeV}$ \cite{nnonline},
around which the amplitude can be expanded as \cite{VanKolck98}
\begin{equation}
\frac{m_N}{4\pi}\, T(k) = \frac{k^2-k_0^2}{k_0^3}
\;\left[z_1 +z_2\frac{k^2-k_0^2}{k_0^2}
+{\cal O}\left(\frac{(k-k_0)^2}{k_0^2}\right) \right]
\label{expansionaroundzero}
\end{equation}
in terms of dimensionless parameters $z_n$,
with $|z_n|={\cal O}(1)$ in the absence of further fine tuning.
One can easily check that $\delta(k)$ behaves linearly around $k=k_0$,
with a slope proportional to $z_1$,
\begin{equation}
\delta(k\sim k_0)=-\frac{2z_1}{k_0}(k-k_0)+\cdots.
\label{phaseshiftslopeatzero}
\end{equation}
From the \textit{Nijm93} phase shifts \cite{nnonline}
we find $z_1 \simeq 0.6$.

It has long been recognized that the anomalously large value
of $|a|$ is a consequence of a fine tuning that places
a virtual bound state very close to threshold, and introduces
an accidental, small scale $\aleph\sim 10$ MeV
corresponding to its binding momentum.
In the standard version of $\slashed{\pi}$EFT,
higher ERE parameters are assumed
to depend on a single higher-energy scale $\tilde{M}_{\mathrm{hi}}$,  
$1/r_0 \sim 1/P_0^{1/3}\sim \cdots= \mathcal{O}(\tilde{M}_{\mathrm{hi}})$.
The PC then organizes the contributions to an observable
characterized by a momentum $Q\sim \aleph$ in an expansion in powers
of $Q/\tilde{M}_{\mathrm{hi}}$, \textit{i.e.} $\tilde{M}_{\mathrm{hi}}$ becomes
the breakdown scale of the theory.
Naively one expects $\tilde{M}_{\mathrm{hi}}\lesssim m_\pi$, but there is some
evidence that $\slashed{\pi}$EFT works
also at larger momenta.
For example, the binding momenta for the ground states
of systems with $A=3,4,6,16$ nucleons
are near 100 MeV, and yet their physics is well described
by the lowest orders of $\slashed{\pi}$EFT (see, for example,
Refs. \cite{Bedaque:1999ve,Platter:2004zs,Stetcu:2006ey,Contessi:2017rww}).
In fact, it has been suggested that
the characteristic scale of $\slashed{\pi}$EFT is set by
these binding momenta through the LO three-nucleon force,
so that $\aleph$ appears only at NLO or higher
\cite{Konig:2015aka,vanKolck:2016}.

Here we propose to accommodate an enlarged range of validity of
$\slashed{\pi}$EFT and the smallness of $1/a$
by changing the standard PC of $\slashed{\pi}$EFT in the $^1S_0$
channel on the basis of the replacements
$\tilde{M}_{\mathrm{hi}} \to M_{\mathrm{lo}}$ and
$\aleph \to M_{\mathrm{lo}}^2/M_{\mathrm{hi}}$.
The phenomenological parameters of the theory
are assumed to scale as
\begin{equation}
1/a = \mathcal{O}\left({M_{\mathrm{lo}}^2}/{M_{\mathrm{hi}}}\right),
\quad
k_0 \sim 1/r_0 \sim 1/P_0^{1/3} \sim \cdots =\mathcal{O}(M_{\mathrm{lo}}),
\label{EREscaling}
\end{equation}
with $M_{\mathrm{hi}}\gg M_{\mathrm{lo}}$.
This assumption will allow us to develop an expansion for an observable
at typical momentum $Q\sim M_{\mathrm{lo}}$ in powers
of $Q/M_{\mathrm{hi}}$.
The usefulness of such an expansion is far from obvious,
but as we show below it seems to give good results.
Our prescription includes the correct position of the amplitude zero at LO,
and moves the virtual state at NLO very close to its empirical
position.
For $Q\sim \aleph$ the NLO amplitude
is similar to that of standard
$\slashed{\pi}$EFT with $\tilde{M}_{\mathrm{hi}}=\mathcal{O}(M_{\mathrm{lo}})$.
The assignment $\aleph \to M_{\mathrm{lo}}^2/M_{\mathrm{hi}}$ is somewhat
arbitrary but motivated by the expectation that
$M_{\mathrm{lo}}\sim 100$ MeV and $M_{\mathrm{hi}}\sim 500$ MeV,
when it holds within a factor of 2 or so.
If $\aleph$ were taken to be smaller, a reasonable description
of observables at momenta $Q\sim \aleph$ would only emerge at higher
orders. Conversely, had we decided to treat $\aleph$ as $M_{\mathrm{lo}}$,
the very-low-energy region would be well reproduced already
at LO, but it would be more difficult to see improvements at NLO.

Quantities in the theory can be organized in powers of
the small expansion parameter $M_{\mathrm{lo}}/M_{\mathrm{hi}}$.
For a generic coupling constant $\mathsf{g}$, we expand formally
\begin{equation}
\mathsf{g}(\Lambda) = \mathsf{g}^{[0]}(\Lambda)+\mathsf{g}^{[1]}(\Lambda)+\cdots,
\label{g}
\end{equation}
where the superscript $^{[\nu]}$ indicates that the coupling appears
at N$^{\nu}$LO.
The ``renormalized'' coupling $\bar{\mathsf{g}}^{[\nu]}$
--- \textit{i.e.} the regulator-independent contribution to the bare (running)
coupling $\mathsf{g}^{[\nu]}\left(\Lambda\right)$ ---
is nominally suppressed by
${\cal O}(M_{\mathrm{lo}}^{\nu}/M_{\mathrm{hi}}^{\nu})$ with respect to
$\bar{\mathsf{g}}^{[0]}$.

Likewise, the amplitude is written
\begin{equation}
T(k;\Lambda) = T^{[0]}(k;\Lambda) + T^{[1]}(k;\Lambda) +\cdots,
\end{equation}
where
\begin{eqnarray}
T^{[\mathrm{0}]}(k;\Lambda) &=&
V^{[0]}(k;\Lambda)\left[1+\frac{m_N}{4\pi}V^{[0]}(k;\Lambda)
\left(ik+\theta_1\Lambda
+ \frac{k^2}{\Lambda}\sum_{n=0}^\infty\theta_{-1-2n}\frac{k^{2n}}{\Lambda^{2n}}\right)
\right]^{-1}, \label{T0}\\
T^{[\mathrm{1}]}(k;\Lambda) &=&
\left(\frac{T^{[\mathrm{0}]}(k;\Lambda)}{V^{[0]}(k;\Lambda)}\right)^2
V^{[\mathrm{1}]}(k;\Lambda), \label{T1}
\end{eqnarray}
{\it etc.}, in terms of
\begin{eqnarray}
V^{[\mathrm{0}]}(k;\Lambda) &=& \frac{4\pi}{m_N}
\sum_{j}\left(\Delta_j^{[\mathrm{0}]}(\Lambda)
+c_j^{[\mathrm{0}]}(\Lambda)\frac{k^2}{m_N}\right)^{-1},
\label{V0}\\
V^{[\mathrm{1}]}(k;\Lambda) &=&  - \frac{4\pi}{m_N}
\sum_j\left(\Delta_j^{[0]}(\Lambda)+c_j^{[0]}(\Lambda) \frac{k^2}{m_N}\right)^{-2}
\left(\Delta_j^{[1]}(\Lambda)+c_j^{[1]}(\Lambda)\frac{k^2}{m_N}\right),
\label{V1}
\end{eqnarray}
{\it etc.}
Neglecting higher-order terms,
the 
%corresponding 
phase shifts at LO, LO+NLO and so on can be written as
\begin{eqnarray}
\delta^{[0]}(k;\Lambda) &=&
-\cot^{-1}\left(\frac{4\pi}{m_Nk}\,\mathrm{Re}\left(\,T^{[0]}(k;\Lambda)\right)^{-1}
\right),
\label{delta0}\\
\delta^{[0+1]}(k;\Lambda)&=&
-\cot^{-1}\left(\frac{4\pi}{m_Nk}
\,\mathrm{Re}\left[\,\left(T^{[0]}(k;\Lambda)\right)^{-1}
\left(1-\frac{T^{[1]}(k;\Lambda)}{T^{[0]}(k;\Lambda)}\right)\,\right]\right),
\label{delta0+1}
\end{eqnarray}
{\it etc.}
At higher orders interactions in the ``$\cdots$'' of Eq. \eqref{L}
appear. We now consider the first two orders of the expansion in detail.

\bigskip
\begin{center}
\subsection{Leading Order}
\label{pionlessLO}
\end{center}

{}From Eq. \eqref{4param} we see that
reproducing the amplitude zero at LO with a shallow pole
requires a minimum of three bare parameters.
Both residual masses,
$\Delta_1(\Lambda)$ and $\Delta_2(\Lambda)$,
must be non-vanishing, otherwise the resulting inverse amplitude
at threshold would be proportional to $\Lambda$, \textit{i.e.}
not properly renormalized.
At the same time, at least one of the kinetic factors,
which we choose to be $c_2(\Lambda)$,
needs to appear at LO, otherwise the amplitude zero could not be reproduced.

Since we attribute in Eq. \eqref{EREscaling}
the smallness of the inverse scattering length to a suppression
by one power of the breakdown scale $M_{\mathrm{hi}}$, we take
\begin{equation}
\frac{1}{a^{[0]}}=0.
\label{inva0}
\end{equation}
In other words, we perform an expansion of the $N\!N$ $^1S_0$ amplitude around
the unitarity limit, as in Refs. \cite{Konig:2015aka,vanKolck:2016}.
One of the dibaryon parameters, which turns out to be $\Delta_2(\Lambda)$,
carries such an effect, so that its observable contribution vanishes at LO.
The regulator-independent parts of the remaining LO parameters, $\Delta_1$
and $c_2$, are assumed to be governed by the scale $M_{\mathrm{lo}}$
\footnote{NDA \cite{Manohar:1983md,Georgi93} gives for a dibaryon-$N\!N$
coupling $g=\mathcal{O}(4\pi/\sqrt{m_N})$, which differs from our convention
\eqref{conv} by a factor of $\sqrt{4\pi}$. Since it is
the combination $g^2/\Delta$
that enters the amplitude, $\Delta$ is expected to be
$\mathcal{O}(M_{\mathrm{hi}}/(4\pi))=\mathcal{O}\left(M_{\mathrm{lo}}\right)$
instead of $\mathcal{O}(M_{\mathrm{hi}})$.}.
In a nutshell,
\begin{equation}
\bar{\Delta}_1^{[0]} = \mathcal{O}\left(M_{\mathrm{lo}}\right), \quad
\frac{\bar{c}_1^{[0]}}{m_N} = 0, \quad
\bar{\Delta}_2^{[0]} = 0, \quad
\frac{\bar{c}_2^{[0]}}{m_N} =\mathcal{O}\left(\frac{1}{M_{\mathrm{lo}}}\right).
\label{LOscaling}
\end{equation}
Because of the vanishing of $c_1^{[0]}$,
eliminating dibaryon-1 via Eq. \eqref{onedibaryon} generates
a momentum-independent contact interaction.
Thus, at LO we obtain
--- except for our additional requirement \eqref{inva0} ---
the $M_{N\!N}\to\infty$ version of the model considered
in Ref. \cite{Kaplan97}, where a dibaryon (our dibaryon-2) is added
to a series of nucleon contact interactions.

In order to relate $\Delta_1^{[0]}(\Lambda)$, $\Delta_2^{[0]}(\Lambda)$, and
$c_2^{[0]}(\Lambda)$ --- our three non-vanishing LO bare parameters ---
to observables, we impose on
\begin{equation}
F(z;\Lambda)
\equiv \mathrm{Re}\left\lbrace\left[\frac{m_N}{4\pi}\,
T^{[\mathrm{0}]}(\sqrt{z};\Lambda)\right]^{-1}\right\rbrace
\end{equation}
three renormalization conditions,
\begin{equation}
F(0;\Lambda) =0, \qquad
\left.
\dfrac{\partial F(z;\Lambda)}{\partial z}\right|_{z=0} = -\frac{r_0}{2},
\qquad
{F}^{-1}(k_0^2;\Lambda)=0.
\label{LOcond}
\end{equation}
The dependence of loops on positive powers of $\Lambda$ is
canceled by that of the bare couplings,
\begin{eqnarray}
\Delta_1^{[0]}(\Lambda)&=& \bar{\Delta}_1^{[0]} - \theta_1\Lambda + \cdots,
\label{Delta1}\\
\Delta_2^{[0]}(\Lambda)&=& \frac{2\theta_1}{r_0^3k_0^2}
\left[\theta_1 \left(r_0\Lambda\right)^2
-\left(\frac{r_0^2k_0^2}{2}+2\theta_1\theta_{-1}\right)r_0\Lambda
+ 4\theta_1\theta_{-1}^2
+ \cdots\right],
\label{Delta2}\\
\frac{c_2^{[0]}(\Lambda)}{m_N}&=& \frac{\bar{c}_2^{[0]}}{m_N}
- \frac{2\theta_1 }{r_0^3k_0^4}
\left[\theta_1 \left(r_0\Lambda\right)^2
-\left(r_0^2k_0^2+2\theta_1\theta_{-1}\right)r_0\Lambda
+ 4\theta_1\theta_{-1}^2
+ \cdots\right],
\label{c2}
\end{eqnarray}
where ``$\cdots$'' stands for terms that become arbitrarily small
for an arbitrarily large cutoff.
Equation \eqref{LOscaling} ensures that
the non-vanishing renormalized couplings,
\begin{equation}
\bar{\Delta}_1^{[0]} = \frac{r_0k_0^2}{2}, \qquad
\frac{\bar{c}_2^{[0]}}{m_N} = -\frac{r_0}{2},
\label{LOren}
\end{equation}
are indeed consistent with Eq. \eqref{EREscaling}.
Apart from a residual cutoff dependence that can be made arbitrarily
small by increasing the cutoff,
the amplitude can now be expressed in terms of the renormalized couplings
or, using Eq. \eqref{LOren}, in terms of $r_0$ and $k_0$:
\begin{equation}
\left[\frac{m_N}{4\pi}\, T^{[0]}(k;\Lambda)\right]^{-1} =
ik
-\frac{r_0}{2}\frac{k^2}{1-k^2/k_0^2}
\left(1+\frac{2\theta_{-1}}{r_0\Lambda}\,\frac{k^2}{k_0^2}
\right)
+\mathcal{O}\left(\frac{k^4}{\Lambda^3}\right).
\label{invT}
\end{equation}
Although the scales and the zero location are different,
Eq. \eqref{invT} is similar to the one \cite{nd}
for $nd$ scattering at very low energies
\footnote{Defining
\begin{equation}
A \equiv \frac{r_0}{2}k_0^2\equiv -R,
\nonumber
\end{equation}
Eq. \eqref{invT} may be rewritten as
\begin{equation}
\left[\frac{m_N}{4\pi}\, T^{[0]}(k;\Lambda)\right]^{-1}=
A+\frac{R}{1-k^2/k_0^2}+ik+\mathcal{O}\left(\frac{k^2}{\Lambda}\right),
\nonumber
\end{equation}
which is a form used in early work on
$nd$ scattering, such as Ref. \cite{VanOers67}.}.

Many interesting consequences can be extracted from Eq. \eqref{invT}.
For momenta below the amplitude zero, our expression reduces
to the unitarity-limit version of the ERE \eqref{ERE} but with
predictions for the higher ERE parameters,
starting with the shape parameter
\begin{equation}
P_0^{[0]}(\Lambda) = \frac{2r_0}{k_0^2}
\left[1+ \frac{2\theta_{-1}}{r_0 \Lambda}
+\mathcal{O}\left(\frac{k_0^2}{r_0 \Lambda^3}\right)\right].
\label{inducedP0}
\end{equation}
Using the cutoff dependence to estimate
the error under the assumption $M_{\textrm{hi}}\sim 500$  MeV,
the LO prediction is $P_0^{[0]} k_0^2/(2r_0)=1.0 \pm 0.3$.
These high ERE parameters are difficult to extract from data.
A careful analysis in Ref. \cite{PB09} obtains
$P_0k_0^2/(2r_0)=1.1$, which
is well within our expected truncation error.
Yet, values obtained for $P_0$ from the phenomenological $np$
potentials \textit{NijmII} and \textit{Reid93} \cite{Nijm93} are
of the same order of magnitude as the value from Ref. \cite{PB09},
but with a \textit{negative} sign \cite{ValderramaArriola04}.

We conjecture that, in contrast to standard $\slashed{\pi}$EFT,
Eq. \eqref{invT} also applies at momenta around the amplitude zero,
with terms which are $\mathcal{O}(M_{\mathrm{lo}})$ and corrections of
$\mathcal{O}(M_{\mathrm{lo}}^2/M_{\mathrm{hi}})$.
Around the amplitude zero, the amplitude is perturbative
\cite{VanKolck98,Lutz00}.
Indeed, a simple Taylor expansion of
Eq. \eqref{invT} gives a perturbative expansion in the region
$|k-k_0|\simle k_0$, \textit{i.e.} an
equation of the form \eqref{expansionaroundzero}
with LO predictions for the coefficients,
\begin{eqnarray}
z_1^{[0]}(\Lambda)&=& \frac{2}{r_0k_0}
\left(1-\frac{2\theta_{-1}}{r_0\Lambda}+\cdots\right),
\label{z1LO}\\
z_2^{[0]}(\Lambda)&=&  -\frac{2}{r_0k_0}
\left[1 +\frac{2i}{r_0k_0}
\left(1-\frac{4\theta_{-1}}{r_0\Lambda}\right)+\cdots\right],
\label{z2LO}
\end{eqnarray}
where the ``$\cdots$'' account for $\mathcal{O}(M_{\mathrm{lo}}^2/\Lambda^2)$.
Numerically, these coefficients are $z_1^{[0]}=0.4\pm 0.1$
and $z_2^{[0]}= - (0.4 \pm 0.1) -i (0.2\pm 0.1)$,
which are indeed of ${\cal O}(1)$.
The former is in fact reasonably close to $z_1\simeq 0.6$
extracted from the phenomenological data.
Note that we could have imposed as a renormalization
condition that $z_1$ had a fixed value
--- the one that best fits the empirical value ---
at any $\Lambda$,
thus trading the information about energy dependence
carried by $r_0$ for that contained in the derivative
of the phase shift at its zero, see Eq. \eqref{phaseshiftslopeatzero}.

Equation \eqref{invT} interpolates between the two regions,
$k\ll k_0$ where the amplitude is non-perturbative
and $|k-k_0|\ll k_0$ where it is perturbative.
Compared to standard $\slashed{\pi}$EFT, it resums not only range corrections
as in Ref. \cite{BeaneSavage}, but also corrections
that give rise to the pole at $k=k_0$. Compared to the expansion around the
amplitude zero \cite{VanKolck98}, it resums the terms that become
large at low energies and give rise to a resonant state at zero energy.
The pole structure of the LO amplitude can be made explicit by
rewriting Eq. \eqref{invT} as
\begin{equation}
\left[\frac{m_N}{4\pi}\, T^{[0]}(k;\Lambda)\right]^{-1} =
\frac{(k - i\kappa_1^{[0]})(k - i\kappa_2^{[0]})(k - i\kappa_3^{[0]})}
{i (k_0^2 - k^2)}
+\mathcal{O}\left(\frac{k^2}{\Lambda}\right),
\label{rewriting}
\end{equation}
with
\begin{equation}
\kappa_1^{[0]}=0,
\quad
\kappa_2^{[0]}=\frac{r_0k_0^2}{4}
\left(1 - \sqrt{1-\left(4/(r_0k_0)\right)^2}\right),
\quad
\kappa_3^{[0]}=\frac{r_0k_0^2}{4}
\left(1 + \sqrt{1-\left(4/(r_0k_0)\right)^2}\right).
\label{kappaj}
\end{equation}
In addition to the amplitude zero, $T^{[0]}(k_0;\Lambda)=0$,
it is apparent that there are three simple poles,
$T^{[0]}(i\kappa_j^{[0]};\infty)\to\infty$,
the nature of which is linked to the sign
of $i\,\mathrm{Res}\,S^{[0]}(i\kappa_j^{[0]})$:

\begin{itemize}

\item The pole at $k=0$
represents a resonant state at threshold, as it induces the vanishing of
$\cot\delta(0)$. Such a pole can be reproduced even with
a momentum-independent contact potential, just as it is done at
LO in standard $\slashed{\pi}$EFT \eqref{Lpiless}
in the unitarity limit.
(Since $i\,\mathrm{Res}\, S^{[0]}(i\kappa_1^{[0]})=0$, this state has
a non-normalizable wavefunction.)

\item The pole at $k=i\kappa_2^{[0]}$, $\kappa_2^{[0]}\simeq 190~\mathrm{MeV}$,
lies on the positive imaginary semiaxis.
However, since $i\,\mathrm{Res}\, S^{[0]}(i\kappa_2^{[0]})<0$,
the condition to produce
a normalizable wavefunction is not satisfied.
The pole at $k=i\kappa_2^{[0]}$
cannot correspond to a bound state, whose wavefunction
has finite support in coordinate space. It is
a \textit{redundant} pole \cite{MA46,MA47}.

\item The pole at $k=i\kappa_3^{[0]}$, $\kappa_3^{[0]}\simeq 600~\mathrm{MeV}$,
lies deep on the positive imaginary semiaxis.
It represents a bound state because
$i\,\mathrm{Res}\, S^{[0]}(i\kappa_3^{[0]})>0$.
Since no such state exists experimentally,
it sets an upper bound on the regime of validity of the EFT,
$M_{\textrm{hi}}\simle \kappa_3^{[0]}$.
\end{itemize}

\begin{figure}[tb]
\includegraphics[scale=.55]{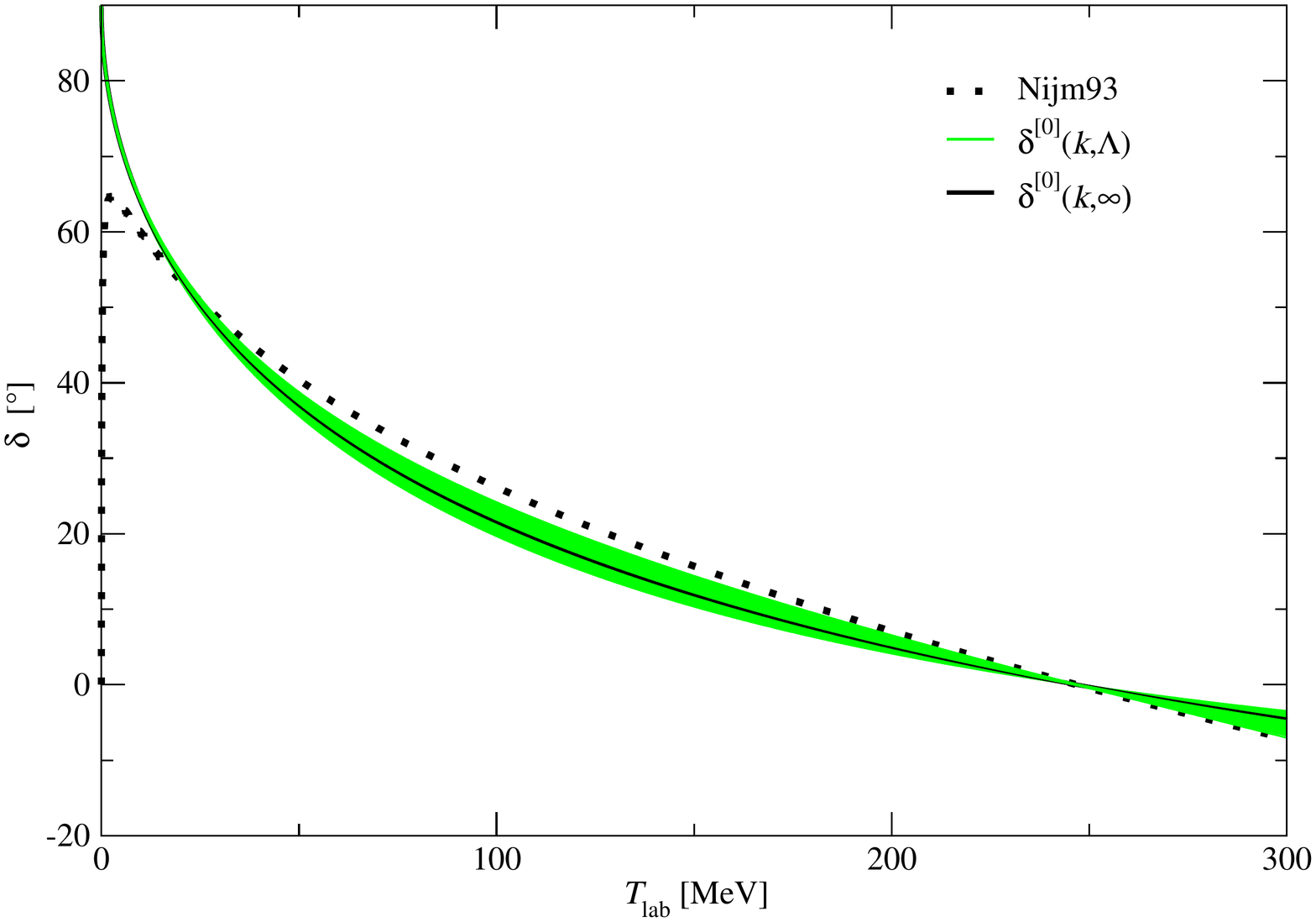}
\caption{
(Color online.) $np$ $^1S_0$ phase shift $\delta$ (in degrees) versus
laboratory energy $T_\mathrm{lab} 
=2k^2/m_N$ (in MeV) for $\slashed{\pi}$EFT at LO
in our new PC.
The (black) solid line shows the analytical result \eqref{invT} with
$\Lambda\to \infty$, while the (green) band around it represents
the evolution of the cutoff from 500 MeV to infinity,
with $\theta_{-1}=\pm 1$.
The (black) squares are the \textit{Nijm93} results \cite{Nijm93,nnonline}.
}
\label{phaseshiftsLOpionless}
\end{figure}

In Fig. \ref{phaseshiftsLOpionless}, we plot the $^1S_0$ phase shifts
\eqref{delta0} from the LO amplitude \eqref{invT}
in comparison with the \textit{Nijm93} results \cite{Nijm93,nnonline}.
As input, we use the empirical values of the effective range and
the position of the amplitude zero.
We display the cutoff band for a generic regulator by taking $\theta_{-1}=\pm 1$
and varying $\Lambda$ from around the breakdown
scale (500 MeV) to infinity
--- as the cutoff increases, our results converge,
as evident in Eq. \eqref{invT}.
This cutoff band provides an estimate of the LO error, except
at low momentum where there is an error that scales with $1/|a|$ instead
of $k$.
The LO phase shifts are in good agreement with empirical values
for most of the low-energy momentum range, except at the very low momenta
where the small but non-vanishing
virtual-state binding energy is noticeable.
A plot of $k \cot\delta$ shows that differences at the amplitude
level are indeed small.
We plot phase shifts to
better display the region around the amplitude zero,
which our PC is designed to capture.
There, while the phase shifts themselves are not too far off
empirical values, the curvature is not well reproduced.
Nevertheless, the agreement is surprisingly good given the absence of
explicit pion fields. In the next section we examine how robust
this agreement is.

\bigskip
\begin{center}
\subsection{Next-to-Leading Order}
\label{pionlessNLO}
\end{center}

As pointed out in Ref. \cite{Long:2012ve}, the leading residual
cutoff dependence of an amplitude, together with the assumption of naturalness,
gives an upper bound on the order of the next correction to that amplitude.
In standard $\slashed{\pi}$EFT, for example, the LO amplitude
has an effective range $r_0 \sim 1/\Lambda$, indicating
that there is an interaction at order no higher than NLO
\cite{vanKolck:1997ut,KSW98bis,KSW98,VanKolck98}
which will produce a physical
effective range $r_0\sim 1/\tilde{M}_{\textrm{hi}}$.
The leading residual cutoff dependence in Eq. \eqref{invT} is
proportional to $k^4$ and of relative order ${\cal O}(M_{\textrm{lo}}/\Lambda)$.
Thus, the NLO interaction must give rise to a contribution
\begin{equation}
P_0^{[1]}(\Lambda) \equiv P_0-P_0^{[0]}(\Lambda)
= {\cal O}\left(\frac{1}{M_{\mathrm{lo}}^{2}M_{\mathrm{hi}}}\right)
\label{P0NLOscaling}
\end{equation}
to the LO shape parameter \eqref{inducedP0}.
This correction requires a higher-derivative operator. Although we could add
a momentum-dependent contact operator,
a simpler, energy-dependent strategy will be implemented here:
we allow for a non-vanishing $c_1^{[1]}$.

In addition, since we are interpreting $\aleph\to M_{\mathrm{lo}}^{2}/M_{\mathrm{hi}}$,
one combination of parameters including $\Delta_2^{[1]}$ enforces
\begin{equation}
\frac{1}{a^{[1]}}= \frac{1}{a}
= {\cal O}\left(\frac{M_{\mathrm{lo}}^{2}}{M_{\mathrm{hi}}}\right).
\label{a1}
\end{equation}
We also introduce corrections in the other two parameters, $c_2^{[1]}$
and $\Delta_1^{[1]}$, in order to keep $r_0$ and $k_0$ unchanged.
Since NLO interactions must all be suppressed
by $M_{\mathrm{hi}}^{-1}$,
\begin{equation}
\bar{\Delta}_1^{[1]} =
\mathcal{O}\left(\frac{M_{\mathrm{lo}}^2}{M_{\mathrm{hi}}}\right),
\quad
\frac{\bar{c}_1^{[1]}}{m_N}=
\mathcal{O}\left(\frac{1}{M_{\mathrm{hi}}}\right),
\quad
\bar{\Delta}_2^{[1]} =
\mathcal{O}\left(\frac{M_{\mathrm{lo}}^2}{M_{\mathrm{hi}}}\right),
\quad
\frac{\bar{c}_2^{[1]}}{m_N}=
\mathcal{O}\left(\frac{1}{M_{\mathrm{hi}}}\right).
\label{NLOscaling}
\end{equation}
This scaling --- together with what was learned at LO ---
is consistent with the imposition of four renormalization conditions on
\begin{equation}
G(z;\Lambda) \equiv
-\mathrm{Re}\left\lbrace\left[\frac{m_N}{4\pi}
T^{[\mathrm{1}]}(\sqrt{z};\Lambda)\right]\left[\frac{m_N}{4\pi}
T^{[\mathrm{0}]}(\sqrt{z};\Lambda)\right]^{-2}\right\rbrace,
\end{equation}
which ensure that $a$, $r_0$, $P_0$, and $k_0$ are fully $\Lambda$ independent
at NLO:
\begin{equation}
G(0;\Lambda)=\frac{1}{a},
\quad
\left.
\frac{\partial G(z;\Lambda)}{\partial z}\right|_{z=0}= 0,
\quad
\left.
\frac{\partial^2G(z;\Lambda)}{\partial z^2}\right|_{z=0} =
-\frac{P_0^{[1]}(\Lambda)}{2},
\quad
G(k_0^2;\Lambda) =0.
\label{NLOcond}
\end{equation}

Defining the renormalized parameters
\begin{eqnarray}
&&\bar{\Delta}_1^{[1]} = \bar{\Delta}_2^{[1]}+\dfrac{3k_0^2}{m_N}\,\bar{c}_1^{[1]},
\qquad\qquad\qquad
\dfrac{\bar{c}_1^{[1]}}{m_N} =
-\dfrac{r_0}{2}\left(1-\dfrac{P_0k_0^2}{2r_0}\right),
\nonumber\\
&&\bar{\Delta}_2^{[1]} = \dfrac{1}{a}
+r_0k_0^2\left(1-\dfrac{P_0k_0^2}{2r_0}\right),
\qquad
\,\,\,\dfrac{\bar{c}_2^{[1]}}{m_N} =
-4\left(\dfrac{\bar{c}_1^{[1]}}{m_N}+\dfrac{\bar{\Delta}_2^{[1]}}{2k_0^2}\right),
\label{NLOren}
\end{eqnarray}
which with Eq. \eqref{NLOscaling} give
Eqs. \eqref{EREscaling} and \eqref{P0NLOscaling},
the cutoff dependence of the bare parameters that guarantees Eq. \eqref{NLOcond}
is
\begin{eqnarray}
\Delta_1^{[\mathrm{1}]}(\Lambda) &=& \bar{\Delta}_1^{[\mathrm{1}]}+\cdots,
\label{deltaDelta1}\\
\frac{c_1^{[\mathrm{1}]}(\Lambda)}{m_N} &=&
\frac{\bar{c}_1^{[\mathrm{1}]}}{m_N}+\cdots,
\label{deltac1}\\
\Delta_2^{[\mathrm{1}]}(\Lambda) &=& \bar{\Delta}_2^{[\mathrm{1}]}
- \frac{\theta_1}{r_0^4}P_0^{[1]}(\Lambda)
\,\Big[\theta_1\left(r_0\Lambda\right)^2
+ \left(r_0^2k_0^2-4\theta_1 \theta_{-1}\right)r_0\Lambda
- 2\theta_{-1}\left(r_0^2k_0^2-6\theta_1\theta_{-1}\right)\Big]
\notag\\
&&
-\, \frac{4\theta_1}{ar_0^2k_0^2} \left(r_0\Lambda-2\theta_{-1}\right)
+\,\cdots, \\
\frac{c_2^{[\mathrm{1}]}(\Lambda)}{m_N} &=& \frac{\bar{c}_2^{[\mathrm{1}]}}{m_N}
+\frac{1}{k_0^2} \left(\bar{\Delta}_2^{[\mathrm{1}]}
- \Delta_2^{[\mathrm{1}]}(\Lambda) \right)
+\,\cdots,
\label{deltac2}
\end{eqnarray}
where the ellipsis account for terms that disappear when we take
$\Lambda\to\infty$.

The NLO contribution to the amplitude, Eq. \eqref{T1}, then satisfies
\begin{equation}
\frac{T^{[1]}(k;\Lambda)}{T^{[0]2}(k;\Lambda)} =
-\frac{m_N}{4\pi}\left[ \frac{1}{a}
+ \frac{r_0}{2k_0^2}\frac{k^4}{1-k^2/k_0^2}
\left(1-\frac{P_0k_0^2}{2r_0}+\frac{2\theta_{-1}}{r_0\Lambda}\right)
+\mathcal{O}\left(\frac{k^4}{\Lambda^3}\right)\right],
\label{T1/T02}
\end{equation}
which is indeed suppressed by one negative power of $M_{\mathrm{hi}}$.
If we resum $T^{[1]}(k;\Lambda)$ while neglecting N$^2$LO corrections,
%and use the LO amplitude \eqref{invT}, 
then
\begin{equation}
\left[\frac{m_N}{4\pi}\left(T^{[0]}(k;\Lambda)
+T^{[1]}(k;\Lambda)\right)\right]^{-1}
= \frac{1}{a}+ik-\frac{r_0}{2}k^2 - \frac{P_0}{4}\frac{k^4}{1-k^2/k_0^2}
+\mathcal{O}\left(\frac{k^6}{k_0^2\Lambda^3}\right).
\label{invTpert}
\end{equation}

Now the ERE \eqref{ERE} is reproduced
for $k< k_0$
with the experimental scattering length and shape parameter.
Additionally, there are predictions for the higher ERE parameters
which are hard to test directly since they are difficult
to extract from data.
The zero at $k_0$ remains unchanged due to
our choice of renormalization condition.
Once expanded around $k=k_0$, the distorted amplitude \eqref{invTpert}
%$T^{[0]}+T^{[1]}$
yields NLO coefficients such as
\begin{eqnarray}
z_1^{[1]}(\Lambda)&=&
z_1^{[0]}(\infty) \left(1-\frac{P_0k_0^2}{2r_0}\right)+\cdots,
\label{z1atNLO}\\
z_2^{[1]}(\Lambda)&=&
z_2^{[0]}(\infty) \left(1-i\, \frac{r_0k_0}{2} \right)^{-1}
\left[2\left(1-\frac{P_0k_0^2}{2r_0}\right)- \frac{i}{ak_0}\right]
+ \cdots, \label{z2atNLO}
\end{eqnarray}
where ``$\cdots$'' stands for $\mathcal{O}(M_{\mathrm{lo}}^3/\Lambda^3)$.
NLO contributions
are of relative $\mathcal{O}(M_{\mathrm{lo}}/M_{\mathrm{hi}})$ with respect
to their LO predictions $z_1^{[0]}$ and $z_2^{[0]}$,
consistently with the residual cutoff dependence
displayed in Eqs. \eqref{z1LO} and \eqref{z2LO}.
Since $z_1^{[0]}(\infty)$ underestimates the slope of the phenomenological phase
shifts around the amplitude zero, a better description of data
requires
$z_1^{[1]}(\infty)> 0$
and thus, according to Eqs. \eqref{inducedP0} and \eqref{z1atNLO},
$P_0\lesssim P_0^{[0]}(\infty)$.
The value given in Ref. \cite{PB09} leads to a small change
$|z_{1,2}^{[1]}(\infty)/z_{1,2}^{[0]}(\infty)|\lesssim 1/10$,
but unfortunately it is $\sim10\%$ {\it larger} than $P_0^{[0]}(\infty)$.
Since Ref. \cite{PB09} provides no error bars
it is difficult to decide whether this is a real problem.
We can reproduce the phenomenological value for $z_1$ with
$P_0^{[1]}(\infty)\simeq -0.6 \, P_0^{[0]}(\infty)$,
which is still compatible with convergence but
not so small a change with respect to LO.
Of course, not all the discrepancy between LO and phenomenology
should come from NLO, but this might be indicative
that something is missing.
We will return to the shape parameter in the next section.

NLO also shifts the LO position of the poles \eqref{kappaj} of the
$S$ matrix. One can obtain these shifts reliably by means of
perturbative tools only for the two shallowest LO poles, finding
in the large-cutoff limit
\begin{equation}
\kappa_1^{[1]}=\frac{1}{a},
\qquad
\kappa_{2}^{[1]}=
-\frac{k_0^2+\kappa_2^{[0]2}}{k_0^2-\kappa_2^{[0]2}}
\left[\frac{1}{a}+
\frac{1}{2}\frac{r_0\kappa_2^{[0]4}}{k_0^2+\kappa_2^{[0]2}}
\left(1-\frac{P_0k_0^2}{2r_0}\right)\right].
\label{shift2}
\end{equation}
We see that, as expected, $|\kappa_1^{[1]}|\sim |\kappa_2^{[1]}|
=\mathcal{O}(M_{\mathrm{lo}}^2/M_{\mathrm{hi}})$,
as long as $\kappa_2^{[0]}=\mathcal{O}(M_{\mathrm{lo}})$.
As a consequence:

\begin{itemize}
\item The shallowest pole is moved from threshold
to $k \simeq - 8i$ MeV, and represents the well-known virtual state.
Its new location almost coincides with the physical one.

\item The redundant pole is moved from $k\simeq190i$ MeV to $k\simeq215i$ MeV,
when the value of $P_0$ given in Ref. \cite{PB09} is used.
This represents a shift of relative size $\sim15\%$ with respect to LO.
Roughly two thirds of
this shift are due to the finiteness of the scattering length,
while the other third corresponds to the NLO correction to the shape parameter.
If we take the value of $P_0$ that gives the slope of
the phenomenological phase shifts at $k_0$, then the shape correction
overcomes the scattering length and the pole moves to $k\simeq 155 i$ MeV,
still a modest shift.
\end{itemize}

\begin{figure}[tb]
\centering
\includegraphics[scale=.55]{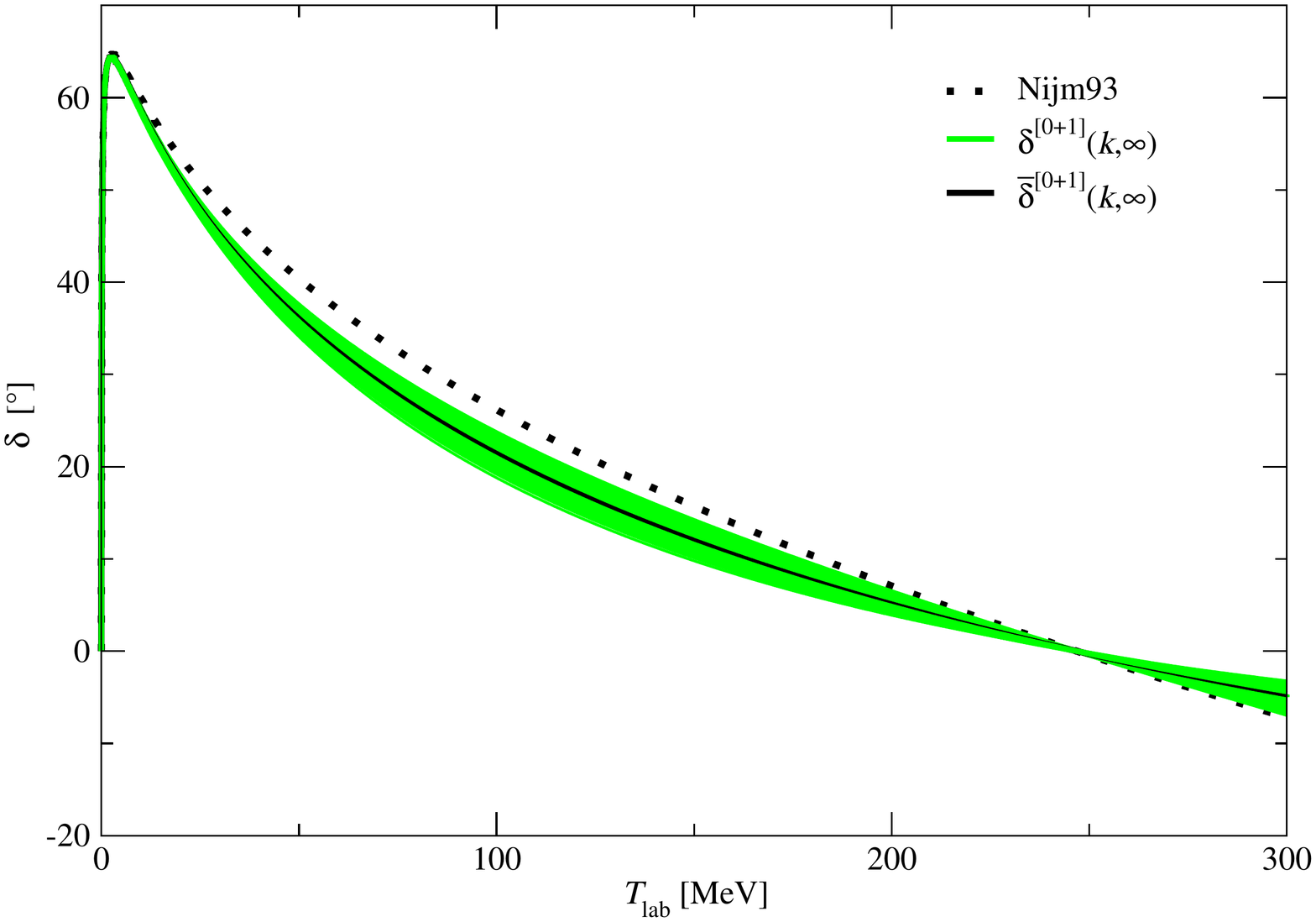}
\caption{(Color online.)
$np$ $^1S_0$ phase shift $\delta$ (in degrees) versus
laboratory energy $T_\mathrm{lab}=2k^2/m_N$ (in MeV) for $\slashed{\pi}$EFT at NLO
in our new PC.
The (black) line shows the analytical result \eqref{T1/T02}
%{invTpert}
with $\Lambda\to \infty$ and the value of the shape parameter from
Ref. \citep{PB09}, while the (green) band around it represents a
$\pm30\%$ variation in this value.
The (black) squares are the \textit{Nijm93} results \cite{Nijm93,nnonline}.
}
\label{phaseshiftsNLOpionless}
\end{figure}

The LO+NLO $^1S_0$ phase shift can now be obtained
from Eqs. \eqref{delta0+1} and \eqref{T1/T02},
see Fig. \ref{phaseshiftsNLOpionless}.
Now, in addition to the empirical values of $r_0$ and $k_0$,
also the values of the scattering length and
the shape parameter from Ref. \cite{PB09} are input.
We show a band corresponding to a variation
of $\pm$30\% around the $P_0$ value of Ref. \cite{PB09}
to account for its (unspecified) error.
Since the cutoff dependence of the
%up-to-NLO
NLO result \eqref{T1/T02}
%{invTpert}
is very quickly convergent ($\sim 1/\Lambda^3$), it has been neglected
in Fig. \ref{phaseshiftsNLOpionless}.
The band thus does not reflect the uncertainty of the NLO
truncation, but of the input.

As expected, the physical value of $a$ greatly improves
the description of the phase shifts at low energies
($k\lesssim50\,\mathrm{MeV}$).
However, at middle energies ($k\sim 100\,\mathrm{MeV}$) this improvement
is much less clear. In particular, as anticipated above, only
for a shape parameter $\sim 30\%$ smaller than in Ref. \cite{PB09}
does $\delta^{[0+1]}(k;\infty)$ get slightly closer to empirical values
than $\delta^{[0]}(k;\infty)$ (see Fig. \ref{phaseshiftsLOpionless}).
Such a change is within the LO error and,
overall, the reproduction of the phase shifts is very good at NLO.
Agreement could be further improved, particularly
around $k_0$, by taking an even smaller
value for the shape parameter --- in particular,
the value that reproduces the phenomenological value for $z_1$.
However, even in that case the
curvature of the resulting phase shifts would
remain different from empirical at middle energies, which
suggests that our expansion is lacking terms at
either LO or NLO.

\bigskip
\begin{center}
\subsection{Resummation and Higher Orders}
\label{pionlessresum}
\end{center}

The choice of identifying the fine-tuning scale $\aleph$ with
$M_{\mathrm{lo}}^2/M_{\mathrm{hi}}$ led to a non-zero scattering length
only at NLO. This assignment is motivated by the numerical values estimated
for these scales.
Alternative choices are possible, leading to
slightly different amplitudes at various orders.
When plotting phase shifts, these differences are amplified.
For example, taking $\aleph$ as $M_{\mathrm{lo}}$ leads to a renormalization
condition where $1/a$ is non-zero already at LO.
In this case our running and renormalized parameters given above
all change by $1/a$ terms. The amplitude itself (or equivalently its
pole positions) changes only slightly, but
in terms of phase shifts there appears to be a large improvement.

Given our previous identification of $\aleph$ with
$M_{\mathrm{lo}}^2/M_{\mathrm{hi}}$,
the alternative procedure just described
amounts to a resummation of higher-order corrections.
Because a bare parameter ($\Delta_2(\Lambda)$) exists already at LO to
ensure proper renormalization, this resummation can be done without harm.
However, because some NLO contributions are shifted to LO,
we see less improvement when going from LO to NLO.
Provided that one has a PC that converges, this is just one of
many ways in which we can make results at one order closer to
phenomenology while remaining within the error of that order.

Regardless of such resummation,
corrections at higher orders are expected to improve the situation further.
The cutoff dependence of Eq. \eqref{invTpert} suggests that
there are no new interactions at next order, N$^2$LO, which
would solely consist of one iteration of the NLO potential.
However, the fact that our pionless phase shifts look too low
in the middle range
represents a significant, systematic lack of attraction between
nucleons at $k\sim m_\pi$.
This could be a reminder to include pions explicitly.
We now consider our expansion with additional pion exchange.

\bigskip
\bigskip
\bigskip
%============================================================
\section{Pionful Theory}
\label{sec:pionful}
%============================================================

We now modify the theory developed in
the previous section to include pion exchange. This is done under
the assumption that the pion mass, the characteristic
inverse strength of OPE,
and the magnitude of the relevant momenta
have similar sizes, not being enhanced or suppressed by
powers of the hard scale:
\begin{equation}
m_\pi \sim M_{N\!N} \sim Q
= \mathcal{O}\left(M_{\mathrm{lo}}\right).
\end{equation}
Such an assumption has been standard in $\chi$EFT since its beginnings
\cite{Weinberg90,Weinberg91}. In the $N\!N$ sector, it underlies
the (non-perturbative) LO character of the OPE interaction,
as well as the suppression of multiple
pion exchanges by powers of $(M_{\mathrm{lo}}/M_{\mathrm{QCD}})^2$.
Moreover, the Coulomb interaction between protons
--- the dominant electromagnetic effect ---
contributes an expansion in $\alpha m_N/M_{\mathrm{lo}} \sim\aleph/M_{\mathrm{lo}}$,
where $\alpha\simeq 1/137$ is the fine-structure constant.
As we took $\aleph = \mathcal{O}(M_{\mathrm{lo}}^2/M_{\mathrm{hi}})$, we should
account for the Coulomb interaction at NLO.
(Other isospin-breaking effects, such as the nucleon mass splitting, are
to be accounted for perturbatively, too.)
Within the $\slashed{\pi}$EFT framework, the (subleading) Coulomb effects
were included in an expansion around the unitarity limit
(without consideration of the amplitude zero) in Ref. \cite{Konig:2015aka}.
Since we anticipate no new features here,
in this first approach we neglect isospin breaking. 
We also ignore the explicit dependence on quark mass, because the 
expansion is already quite complicated at a fixed value of $m_\pi^2$.

Pions are introduced in the usual way, by demanding that the most general
effective Lagrangian transforms under chiral symmetry as does
the QCD Lagrangian written in terms of quarks and gluons.
(For reviews and references, see Refs.
\cite{VanKolckBedaque02,ModTheNucFor,Machleidt'n'Entem}.)
In the particular case of one dibaryon field, this was done in
Ref. \cite{Kaplan97}. The extension to the two dibaryons
of the previous section is straightforward.
If $\vec{\pi}$ is the pion isotriplet,
the effective Lagrangian reads
\begin{align}
\mathcal{L}_\chi^{(2\phi)} =&
\,\frac{1}{2} \left(\partial_\mu\vec{\pi}\cdot\partial^\mu\vec{\pi}
- m_\pi^2{\vec{\pi}}^2\right)
+ N^\dagger\left[i\partial_0 + \frac{\bm{\nabla}^2}{2m_N}
-\frac{g_A}{2f_\pi}\vec{\tau}\cdot
\left(\bm{\sigma}\cdot\bm{\nabla}\vec{\pi}\right)
\right]N \notag
\\
& + \sum_{j=1,2} \left\lbrace{\vec{\phi}}_j^{\,\dagger}
\cdot\left[\Delta_j+c_j \left(i\partial_0 + \frac{\bm{\nabla}^2}{4m_N}\right)
\right]
\vec{\phi}_j - \sqrt{\frac{4\pi}{m_N}}
\left({\vec{\phi}_j}^{\,\dagger} \cdot N^T \vec{P}_{^1S_0}N + \mathrm{H.c.}
\right)\right\rbrace +\cdots,
\label{Lchi}
\end{align}
in the same notation as Eq. \eqref{L}.
The omitted terms, which include chiral partners of the terms
shown explicitly, are not needed up to NLO.

Inspired by the pionless theory, we 
use for the pionful case the 
same dibaryon arrangement of short-range potentials as in 
Sec.~\ref{sec:pionless}. 
Adding the long-range, 
spin-singlet projection of OPE, the LO potential is
%we have the following LO potential:
\begin{eqnarray}
\frac{m_N}{4\pi}V^{[0]}(\bm{p}',\bm{p},k;\Lambda) &=&
-\frac{m_\pi^2}{M_{N\!N}}\frac{1}{(\bm{p}'-\bm{p})^2+m_\pi^2}
+\frac{1}{\Delta_1^{[0]}(\Lambda)}
+ \frac{1}{\Delta_2^{[0]}(\Lambda)+c_2^{[0]}(\Lambda)\, k^2/m_N}
\nonumber\\
&\equiv& \frac{m_N}{4\pi}
\left(V^{[0]}_{\mathrm{L}}(\bm{p}',\bm{p})
+V^{[0]}_{\mathrm{S}}(k;\Lambda)\right),
\label{V0pi}
\end{eqnarray}
where $\bm{p}$ ($\bm{p}'$) is the relative incoming (outgoing) momentum
and the inverse OPE strength is defined as \cite{KSW98bis,KSW98}
\begin{equation}
M_{N\!N}\equiv \frac{16\pi f_\pi^2}{g_A^2m_N}\approx 290\,\mathrm{MeV}.
\end{equation}
The momentum-independent, contact piece of OPE has been
absorbed in the short-range potential $V^{[0]}_{\mathrm{S}}$ through
the redefinition
\begin{equation}
\left(1/\Delta_1^{[0]}(\Lambda)+1/M_{N\!N}\right)^{-1} \to \Delta_1^{[0]}(\Lambda).
\end{equation}
The long-range part of OPE is the Yukawa potential
represented by $V^{[0]}_{\mathrm{L}}$.
Integrating out dibaryon-1 we obtain the potential considered previously in
Ref. \cite{Kaplan97}.
Since two-pion exchange (TPE) enters only at N$^2$LO and higher
\cite{Ordonez:1992xp,Ordonez:1995rz},
at NLO the interaction is entirely short-ranged,
\begin{eqnarray}
\frac{m_N}{4\pi}V^{[1]}(k;\Lambda)
&=&
-\,\frac{\Delta_1^{[1]}(\Lambda)+c_1^{[1]}(\Lambda) \, k^2/m_N}
{\Delta_1^{[0]2}(\Lambda)}
-\frac{\Delta_2^{[1]}(\Lambda)+c_2^{[1]}(\Lambda)\, k^2/m_N}
{\left(\Delta_2^{[0]}(\Lambda)+c_2^{[0]}(\Lambda)\, k^2/m_N\right)^2}.
\label{V1pi}
\end{eqnarray}
In the limit $\Delta_2^{[0]}\to\infty$ the potential is
an energy-dependent version of the momentum-dependent LO+NLO interaction of
Ref. \cite{Long:2012ve},
while the interaction of Ref. \cite{Long13} emerges
in the limit $\Delta_1^{[0]}\to\infty$.

Because OPE cannot be iterated analytically to all orders, we can no longer 
show explicitly that the amplitude has a zero at LO or that the amplitude is 
RG invariant. However, these two important features of the pionless theory are 
expected to be retained by the pionful theory on the basis that the strength 
of OPE is known to be numerically moderate in spin-singlet channels and 
that $V^{[0]}_{\mathrm{L}}$ is non-singular. Moreover, we continue to use 
the scalings 
%Eqs. 
\eqref{LOscaling} and \eqref{NLOscaling}.
%in the pionful theory.
Below we confirm through numerical calculations that the EFT obeying such a 
PC indeed has an amplitude zero and preserves RG invariance.

\bigskip
\begin{center}
\subsection{Leading Order}
\label{pionfulLO}
\end{center}

The off-shell LO amplitude is found from the LO potential \eqref{V0pi}
by solving the LS equation
\begin{equation}
T^{[0]}(\bm{p}',{\bm{p}}, k;\Lambda)
= V^{[0]}(\bm{p}',{\bm{p}}, k;\Lambda)
- m_N\int \frac{\mathrm{d}^3q}{(2\pi)^3}
\frac{f_{\mathrm{R}}(q/\Lambda)}{q^2-k^2-i\epsilon}\,
V^{[0]}(\bm{p}',\bm{q},k;\Lambda) \, T^{[0]}(\bm{q},\bm{p},k;\Lambda),
\label{T0chi}
\end{equation}
with $f_{\mathrm{R}}(q/\Lambda)$ a non-local regulator function \eqref{regulator}.
Defining the Yukawa amplitude,
\begin{equation}
T_{\mathrm{L}}^{[0]}(\bm{p}',{\bm{p}},k;\Lambda)
= V_{\mathrm{L}}^{[0]}(\bm{p}',{\bm{p}})
- m_N\int \frac{\mathrm{d}^3q}{(2\pi)^3}
\frac{f_{\mathrm{R}}(q/\Lambda)}{q^2-k^2-i\epsilon}
\,V_{\mathrm{L}}^{[0]}(\bm{p}',\bm{q})\, T_{\mathrm{L}}^{[0]}(\bm{q},\bm{p}, k;\Lambda),
\label{TL}
\end{equation}
the Yukawa-dressing of the incoming/outgoing $N\!N$ states,
\begin{equation}
\chi_{\mathrm{L}}^{[0]}(\bm{p}, k;\Lambda) = 1
- m_N\int\frac{\mathrm{d}^3q}{(2\pi)^3}
\frac{f_{\mathrm{R}}(q/\Lambda)}{q^2-k^2-i\epsilon}\,
T_{\mathrm{L}}^{[0]}(\bm{p},\bm{q},k;\Lambda),
\label{chi}
\end{equation}
and the resummation of $N\!N$ bubbles with iterated OPE in the middle,
\begin{equation}
\mathcal{I}_{\mathrm{L}}^{[0]}(k;\Lambda) =
4\pi\int\frac{\mathrm{d}^3q}{(2\pi)^3}
\frac{f_{\mathrm{R}}(q/\Lambda)}{q^2-k^2-i\epsilon}\,
\chi_{\mathrm{L}}^{[0]}(\bm{q},k;\Lambda),
\label{IL}
\end{equation}
Eq. \eqref{T0chi} can be rewritten as \cite{KSW96}
\begin{equation}
\left[\frac{m_N}{4\pi}
\left(T^{[0]}({\bm{p}}',\bm{p},k;\Lambda)
- T_{\mathrm{L}}^{[0]}({\bm{p}}',\bm{p},k;\Lambda)\right)\right]^{-1}
=\frac{\left[m_N V_{\mathrm{S}}^{[0]}(k;\Lambda)/(4\pi)\right]^{-1}
+ \mathcal{I}_{\mathrm{L}}^{[0]}(k;\Lambda)}
{\chi_{\mathrm{L}}^{[0]}({\bm{p}}',k;\Lambda)\;
\chi_{\mathrm{L}}^{[0]}(\bm{p},k;\Lambda)}.
\label{offs}
\end{equation}
This is the generalization of Eq. \eqref{D} for LO in the presence
of pions. Because $V_{\mathrm{L}}^{[0]}$ is regular,
the cutoff dependence of the integrals $T_{\mathrm{L}}^{[0]}$ and
$\chi_{\mathrm{L}}^{[0]}$ is only residual,
\textit{i.e.} suppressed by powers of $\Lambda$.
In contrast, just as the $\mathcal{I}_{0}$ in  Eq. \eqref{D},
$\mathcal{I}_{\mathrm{L}}^{[0]}$ has a linear cutoff dependence due to the
singularity of $V_{\mathrm{S}}^{[0]}$.
Additionally, it exhibits a logarithmic dependence
$\sim (m_\pi^2/M_{N\!N}) \, \ln\Lambda$ \cite{KSW96}
arising from the interference between $V_{\mathrm{L}}^{[0]}$ and $V_{\mathrm{S}}^{[0]}$.
This cutoff dependence is at the root of
one of the shortcomings of NDA in the $N\!N$ system.
%Here it can also be absorbed
%in the dibaryon parameters, which then become quark-mass dependent.
%Such a dependence is examined in App. \ref{AppA}.
%The chiral transformation properties of the dibaryons'
%residual mass and kinetic terms are somewhat complicated.
%For the one-dibaryon theory, they can be found in Ref. \cite{SotoTarrus08}.

Compared to Refs. \cite{KSW96,Long:2012ve,Long13},
our $V_{\mathrm{S}}^{[0]}$ has a different $k$ dependence.
As in the previous section,
two dibaryon parameters are needed to describe
the zero of the amplitude and its energy dependence near threshold,
while the third parameter ensures the fine tuning that leads to a
large scattering length.
These three parameters are sufficient for renormalization,
leaving behind only residual cutoff dependence.
Our LO amplitude is analogous to that of Ref. \cite{Lutz00}, which results from
the unitarization of an expansion around the amplitude zero.

Taking the sharp-cutoff function $f_{\mathrm{R}}(x)=\theta(1-x)$,
we solve numerically the
$S$-wave projection of Eq. \eqref{T0chi},
as done in, \textit{e.g.}, Refs. \cite{Long:2012ve,YangPhillips}.
In order to determine the values of the three
bare parameters at a given cutoff,
three cutoff-independent conditions on the amplitude are needed.
We choose them to be the same as in the
previous section, \textit{i.e.}
\begin{itemize}
\item
unitarity limit, $1/a^{[0]}=0$;
\item
physical effective range, $r_0=2.7\,\mathrm{fm}$;
\item
physical amplitude zero, $k_0=340.4\,\mathrm{MeV}$.
\end{itemize}
The values of $\Delta_1^{[0]}(\Lambda)$, $\Delta_2^{[0]}(\Lambda)$, and
$c_2^{[0]}(\Lambda)$ in our numerical calculations must
be very well tuned in order to reproduce the required values of $1/a^{[0]}$,
$r_0$, and $k_0$ within a given accuracy. The need for such a tuning becomes
more and more noticeable as $\Lambda$ is increased \cite{YangPhillips}.
But the resulting phase shift changes dramatically depending on whether
$1/a^{[0]}$ is very small and negative
(for a shallow virtual state) or very small and positive
(as it would correspond to a bound state close to threshold).
Thus, in order to facilitate the numerical
solution of the LS equation, we kept the scattering length large and negative,
$a^{[0]}=-600$ fm.
The difference with the unitarity limit cannot be seen in the results
presented below.

\begin{figure}[tb]
\centering
\includegraphics[scale=.55]{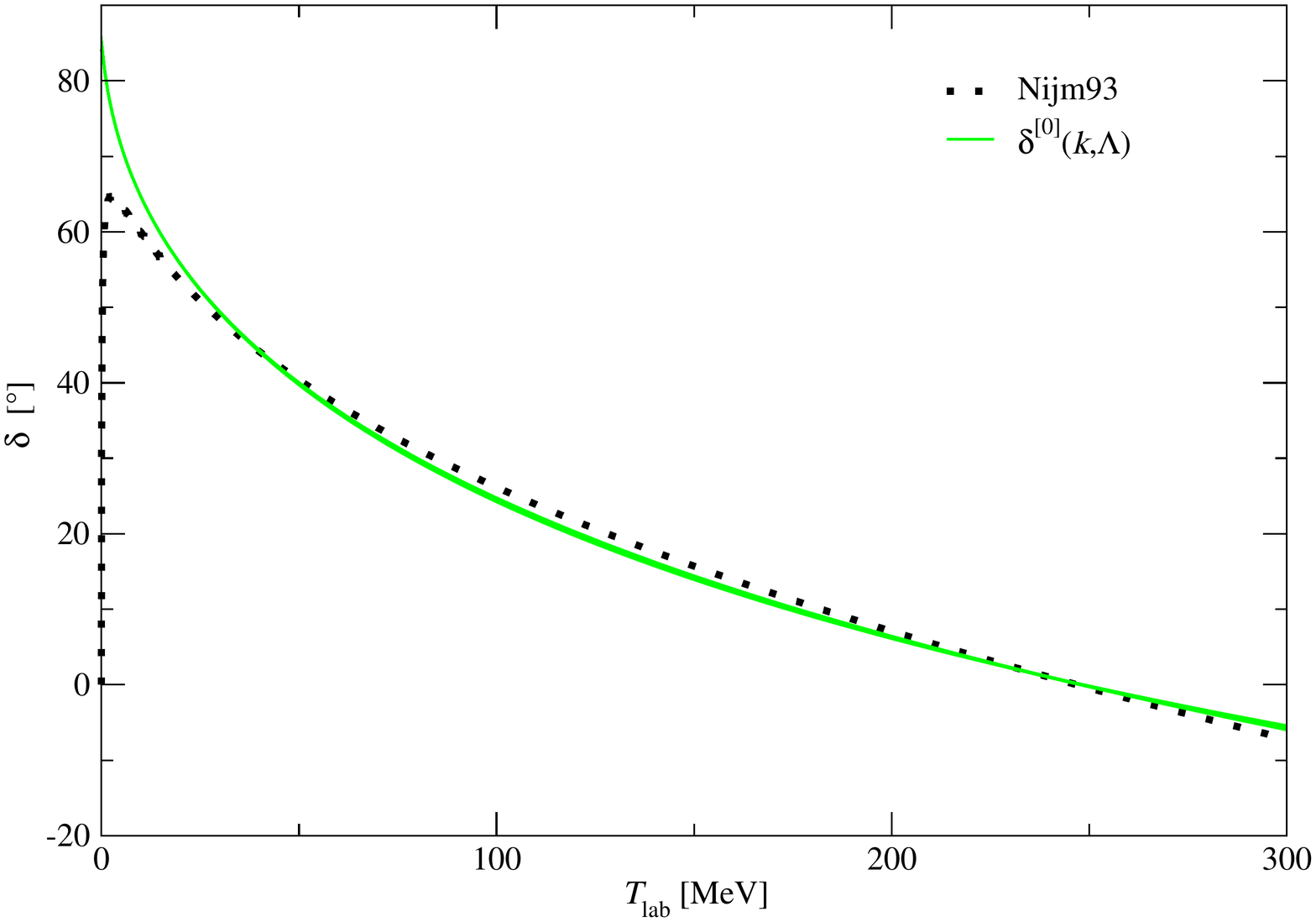}
\caption{
(Color online.) $np$ $^1S_0$ phase shift $\delta$ (in degrees) versus
laboratory energy $T_\mathrm{lab}=2k^2/m_N$ (in MeV) for $\chi$EFT at LO
in our new PC.
The narrow (green) band represents the evolution of the sharp cutoff
from 600 MeV to 2 GeV.
The (black) squares are the \textit{Nijm93} results \cite{Nijm93,nnonline}.
}
\label{chiraldelta}
\end{figure}

The LO pionful phase shift is obtained from the on-shell, $S$-wave-projected
$T$ matrix in the usual way \eqref{delta0}.
The result, presented in Fig. \ref{chiraldelta},
shows little cutoff dependence,
even though the cutoff parameter is varied from 600 MeV to 2 GeV.
It is likely that a more realistic estimate of the systematic error coming
from the EFT truncation is obtained via the variation of the input inverse
scattering length between its physical value and zero. We will come back to
such an estimate later, when we resum finite-$a$ effects.
In any case, comparing with Fig. \ref{phaseshiftsLOpionless}
we confirm that pions help us achieve a better description of phase shifts
between threshold and the amplitude zero.

\begin{figure}[tb]
\includegraphics[scale=.55]{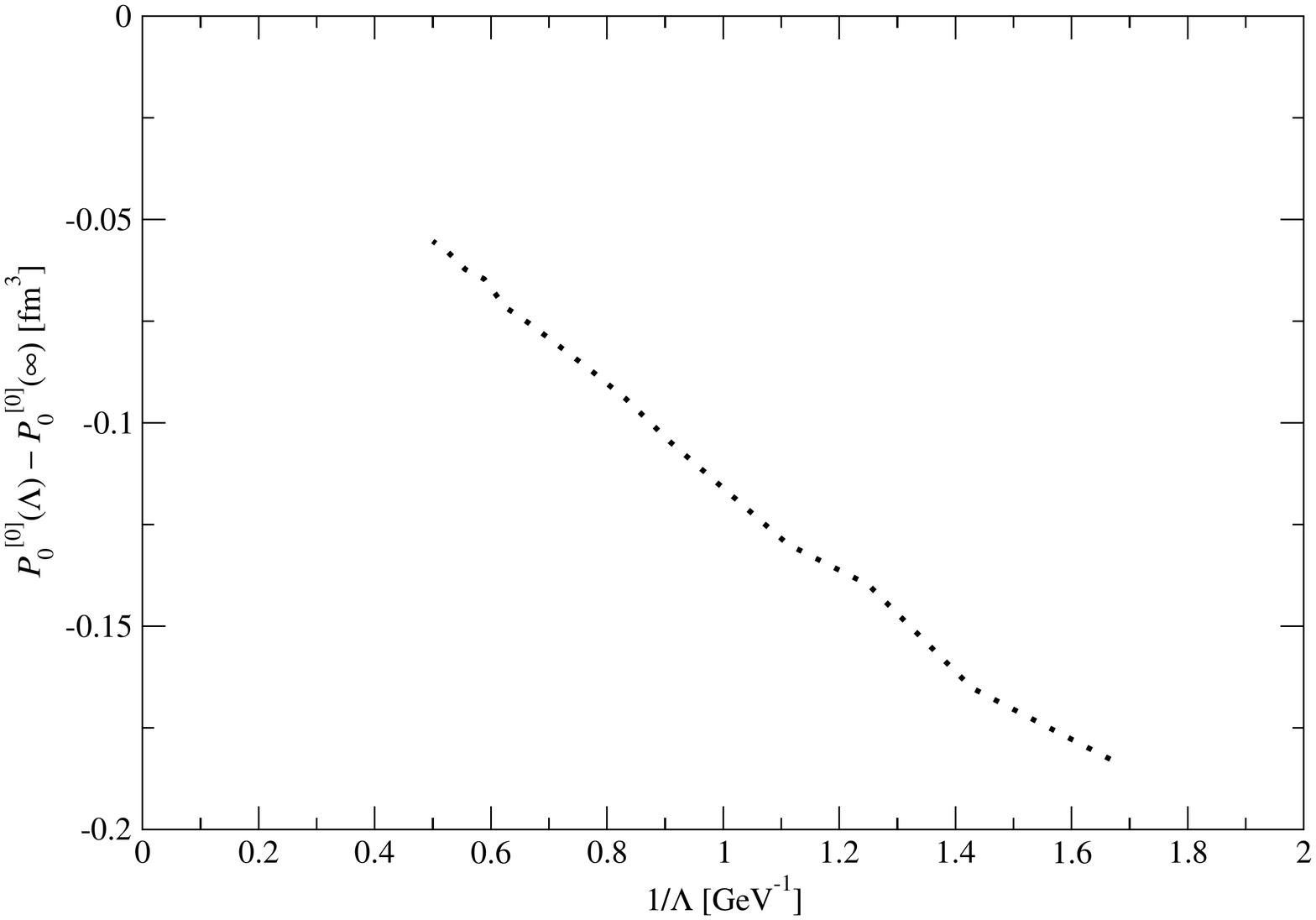}
\caption{$np$ $^1S_0$ shape parameter $P_{0}^{[0]}(\Lambda)$ (in fm$^{3}$)
versus inverse cutoff $1/\Lambda$ (in GeV$^{-1}$) for $\chi$EFT
at LO in our new PC.
}
\label{resdep}
\end{figure}

From the results in Fig. \ref{chiraldelta} we can obtain numerical predictions
for parameters appearing in the ERE and in the expansion
around the amplitude zero.
As an example, we extract the LO shape parameter $P_{0}^{[0]}(\Lambda)$
using our low-energy results and the unitarity-limit version of the ERE
\eqref{ERE} truncated at the level of the shape parameter.
Results are shown in Fig. \ref{resdep}.
For $\Lambda$ large enough, we find
\begin{equation}
P_0^{[0]}(\Lambda) \approx P_0^{[0]}(\infty)
\left(1+ \frac{Q_{P}}{\Lambda}\right),
\end{equation}
with
$P_0^{[0]}(\infty) \approx -1.0\,\mathrm{fm^3}$
and $Q_P \approx 100 \,\mathrm{MeV}$.
Unlike the result for the shape parameter given in Ref. \cite{PB09},
$P_{0}^{[0]}(\infty)$ is \textit{negative}, being reasonably close
to $P_0=-1.9\,\mathrm{fm^3}$ --- the value extracted in
Ref. \cite{ValderramaArriola04} from the \textit{NijmII} fit \cite{Nijm93}.
The large change in the prediction for
$P_{0}^{[0]}(\infty)$ compared
to the corresponding pionless result \eqref{inducedP0}
is confirmation of the importance of pions at LO.

\bigskip
\begin{center}
\subsection{Next-to-Leading Order}
\label{pionfulNLO}
\end{center}

As before, we can infer the short-range contributions at NLO from
the residual cutoff dependence of the amplitude.
Figure \ref{resdep} shows that the cutoff dependence of
$P_{0}^{[0]}(\Lambda)$ is proportional to $1/\Lambda$,
with $Q_P={\cal O}(M_\textrm{lo})$ as expected.
%A heuristic argument in support of this numerical result is presented in
%App. \ref{AppA}.
Just as in the pionless case, this behavior implies that
at least one extra short-range parameter needs to be included at NLO.
This is represented by the NLO potential $V^{[1]}$, Eq. \eqref{V1pi}.

Treating $V^{[1]}$ in distorted-wave perturbation theory,
we obtain a separable NLO amplitude,
\begin{equation}
T^{[1]}(\bm{p}',{\bm{p}}, k;\Lambda)
= \chi^{[0]}(\bm{p}', k;\Lambda)\,V^{[1]}(k;\Lambda)\,
\chi^{[0]}(\bm{p}, k;\Lambda),
\label{T1chi}
\end{equation}
where
\begin{equation}
\chi^{[0]}(\bm{p}, k;\Lambda) = 1
- m_N\int\frac{\mathrm{d}^3q}{(2\pi)^3}
\frac{f_{\mathrm{R}}(q/\Lambda)}{q^2-k^2-i\epsilon}\,
T^{[0]}(\bm{p},\bm{q},k;\Lambda),
\label{chi0}
\end{equation}
is defined in terms of the {\it full} LO amplitude in analogy with
Eq. \eqref{chi} for the long-range LO amplitude.
As in the pionless case, we obtain the pionful LO+NLO phase shift
from Eq. \eqref{delta0+1}.

The dibaryon parameters are fixed in virtue of four cutoff-independent
conditions, which we choose to be the values of the
{\it Nijm93} phase shifts \cite {nnonline} at four different momenta:

\begin{itemize}
\item
$\delta^{[0+1]}(20.0\,\mathrm{MeV};\Lambda)=61.1^\circ$;
\item
$\delta^{[0+1]}(40.5\,\mathrm{MeV};\Lambda)=64.5^\circ$;
\item
$\delta^{[0+1]}(237.4\,\mathrm{MeV};\Lambda)=21.7^\circ$;
\item
$\delta^{[0+1]}(340.4\,\mathrm{MeV};\Lambda)=0^\circ$.
\end{itemize}

\begin{figure}[tb]
\centering
\includegraphics[scale=.55]{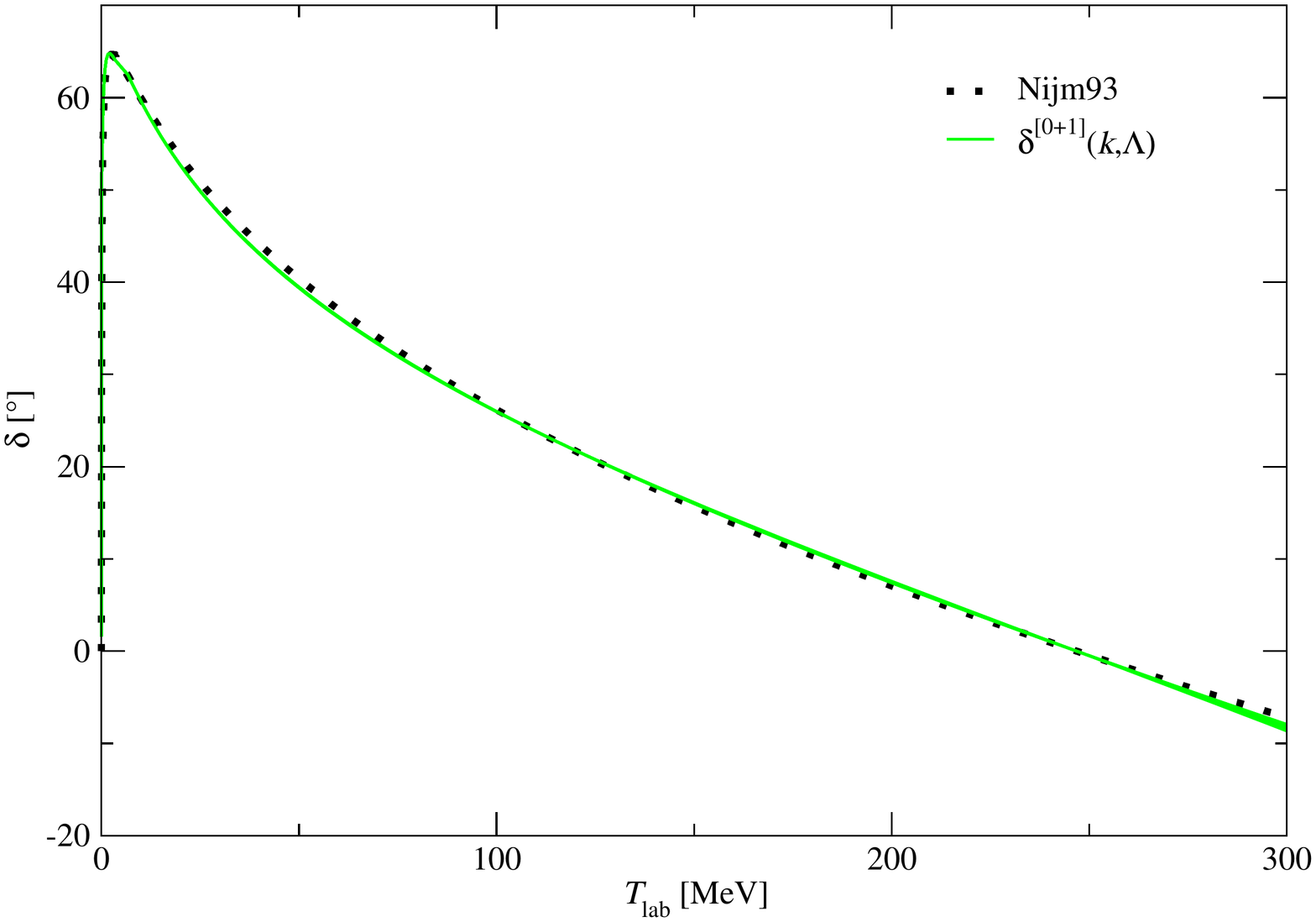}
\caption{
(Color online.) $np$ $^1S_0$ phase shift $\delta$ (in degrees) versus
laboratory energy $T_\mathrm{lab}=2k^2/m_N$ (in MeV) for $\chi$EFT at NLO
in our new PC.
The narrow (green) band represents the evolution of the  cutoff
from 600 MeV to 2 GeV.
The (black) squares are the \textit{Nijm93} results \cite{Nijm93,nnonline}.
}
\label{chiraldeltaNLO}
\end{figure}

The LO+NLO phase shifts are shown in Fig. \ref{chiraldeltaNLO}. 
The narrow band when the cutoff is varied from 600 MeV to 2 GeV confirms
that, as in Fig. \ref{chiraldelta}, very quick cutoff convergence takes
place. The LO+NLO prediction almost lies on the \textit{Nijm93} curve,
which means that now the description of the empirical phase shifts throughout
the whole elastic range $0\lesssim k\lesssim \sqrt{m_Nm_\pi}$ is much better
than at LO.
%(Fig. \ref{chiraldelta}).
Indeed, the improvement is clear not only in the very-low momentum regime
(which had been expected considering that now we relaxed the unitarity-limit
condition), but --- more importantly from the $\chi$EFT point of view ---
also for momenta $k\sim m_\pi$. Comparison with the pionless result at NLO
(Fig. \ref{phaseshiftsNLOpionless}) confirms that adding OPE significantly
improves predictions in this momentum range.

\bigskip
\begin{center}
\subsection{Resummation and Higher Orders}
\label{pionfulresum}
\end{center}

Despite the systematic improvement and good description of data at NLO,
one might be distressed by the unusual appearance of our LO phase shift
(Fig. \ref{chiraldelta})
at low momentum. Within potential models
--- whether purely phenomenological or based on Weinberg's prescription ---
it is traditional to attempt to describe all regions
below some arbitrary momentum on the same footing.

As emphasized earlier, plotting phase shifts
is misleading when it comes to errors in the amplitude,
which is the observable the PC is designed for.
A plot of $k\cot \delta$
shows that only a small amount of physics is missed at LO even at low energies.
Our strategy is a consequence of the fact that the PC assumes momenta
$Q \sim M_\textrm{lo}$, and it is in principle only in
this region
that we expect systematic improvement order
by order. The higher the momentum, the smaller the relative improvement
with order, till we reach $M_\textrm{hi}$ and the EFT stops working.
In the other direction, that of smaller momenta, the $\chi$EFT PC
%\textcolor{red}{which is designed to emphasize at $Q\sim m_{\pi}$ may}
%we already said that a sentence or two ago... 
may no longer capture the relative importance of interactions
properly. A simple example is pion-nucleon scattering in Chiral
Perturbation Theory,
where sufficiently close to threshold the LO $P$-wave interaction
(stemming from the axial-vector coupling in Eq. \eqref{Lchi})
is smaller than NLO corrections to the $S$ wave.
Therefore the region of momenta much below the pion mass
is {\it not} where one wants to judge the convergence of $\chi$EFT.

%In addition, 
However, it might be of practical interest to
improve the description near threshold already at LO.
As in $\slashed{\pi}$EFT, we can choose to reproduce the empirical
value of a phase shift in the very low-momentum region
--- thus accounting for non-vanishing $1/a$ already at LO ---
without doing damage to renormalization.
As is the case with any other choice of data to fit, the difference
with respect to what we have done earlier in this section is of NLO:
we are just resumming some higher-order contributions into LO.

\begin{figure}[tb]
\centering
\includegraphics[scale=.55]{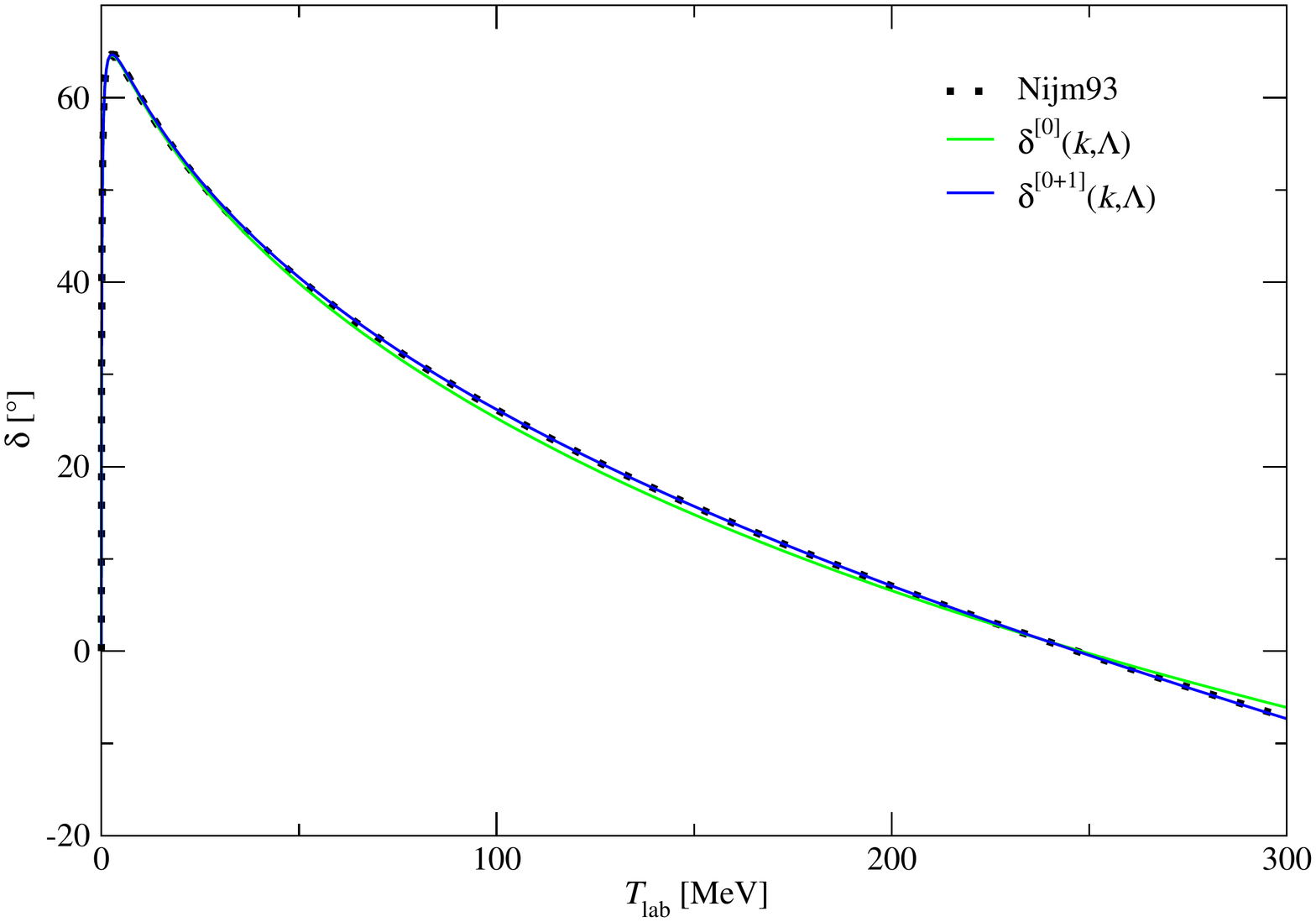}
\caption{
(Color online.) $np$ $^1S_0$ phase shift $\delta$ (in degrees) versus
laboratory energy $T_\mathrm{lab}=2k^2/m_N$ (in MeV) for $\chi$EFT at LO
and NLO in our new PC from an alternative fitting protocol.
The (green) light and (blue) dark bands represent, respectively,
LO and LO+NLO under a
 cutoff variation from $600\,\mathrm{MeV}$ to
$2\,\mathrm{GeV}$.
The (black) squares are the \textit{Nijm93} results \cite{Nijm93,nnonline}.
}
\label{chiraldeltaalternativefacts}
\end{figure}

As an example, in Fig. \eqref{chiraldeltaalternativefacts}
we show LO and LO+NLO results with an alternative fitting protocol.
In the renormalization conditions at LO we replace
the unitarity limit of our original fit with
the physical scattering length, that is, we
impose the following cutoff-independent conditions:
\begin{itemize}
\item
$a=-23.7\,\mathrm{fm}$;
\item
$r_0=2.7\,\mathrm{fm}$;
\item
$k_0=340.4\,\mathrm{MeV}$.
\end{itemize}
Likewise, at NLO we substitute the lowest {\it Nijm93} phase shift of our
earlier fit with the physical scattering length:
\begin{itemize}
\item
$a=-23.7\,\mathrm{fm}$;
\item
$\delta^{[0+1]}(40.5\,\mathrm{MeV})=64.5^\circ$;
\item
$\delta^{[0+1]}(237.4\,\mathrm{MeV})=21.7^\circ$;
\item
$\delta^{[0+1]}(340.4\,\mathrm{MeV})=0^\circ$.
\end{itemize}
As before we vary the cutoff from $600\,\mathrm{MeV}$ to
$2\,\mathrm{GeV}$,
but the $\Lambda$ convergence of the phase shifts
is so quick that the cutoff bands cannot be resolved in our plot.
The improved description of the very low-energy region at LO compared to
that seen in Fig. \ref{chiraldelta}, which is entirely
due to the resummation of the finite scattering length, is evident.
The predicted LO+NLO phase shifts virtually lie on the
the \textit{Nijm93} curve, and this fit is even more phenomenologically
successful than the original LO+NLO shown in Fig. \ref{chiraldeltaNLO}.
The relatively small improvement over the alternative LO curve
%\textcolor{red}{might be a} 
%what else can it be?
is consequence of the resummation of higher-order contributions
into LO. The small difference between alternative
and original LO+NLO curves attests to the fine-tuning of the $^1S_0$ channel,
\textit{i.e.} to the smallness of $1/a$ effects.

Given the importance of OPE, one expects potentially large
changes in the  position of the poles of $T^{[0]}$ in $\chi$EFT with respect to
the $\slashed{\pi}$EFT result \eqref{kappaj}. Yet,
the virtual state near threshold (at $k\simeq i/a$) is guaranteed by
construction, since
%(Eq. \eqref{ERE})
\begin{equation}
\frac{m_N}{4\pi}T^{[0]}(k;\Lambda) \stackrel{k\to0}{\simeq}
\left(\frac{1}{a}+ik\right)^{-1}.
\end{equation}
Using the
technique described in Ref. \cite{YangPhillips}, one may obtain numerically
the positions of the other two poles.
The redundant pole seems to become deeper and deeper when the
cutoff $\Lambda$ is increased.
This is consistent with the point of view that the redundant pole
accounts in $\slashed{\pi}$EFT for the neglected left-hand cut due to OPE.
In contrast, the binding energy of the deep bound state oscillates
with $\Lambda$, but we always find it to be $\gtrsim 200~\mathrm{MeV}$,
which corresponds to a binding momentum $\gtrsim 450~\mathrm{MeV}$.
This is, again, an estimate for the breakdown scale $M_\textrm{hi}$.

One might worry that the LO+NLO result shown in
Fig. \ref{chiraldeltaalternativefacts}
is so good that higher orders could destroy agreement with the empirical
phase shifts and undermine the consistency of our EFT expansion.
At N$^2$LO and N$^3$LO there are several contributions to account for:
TPE and the associated N$^2$LO counterterms \cite{Ordonez:1992xp,Ordonez:1995rz}
in first-order distorted-wave perturbation theory,
as well as NLO interactions in second- and third-order distorted-wave
perturbation theory.
At these higher orders it might be convenient to
use the perturbation techniques of Ref. \cite{Vanasse:2013sda}
or to devise further resummation of NLO interactions.

To investigate the potential effects of higher-order corrections
we have performed an incomplete
N$^2$LO calculation where the long-range component of the N$^2$LO TPE
potential was
included in first-order distorted-wave perturbation theory,
following the analogous calculation in Ref. \cite{Long:2012ve}. Since
the short-range component of this potential can be absorbed in
Eq. \eqref{V1pi}, there are no new short-range parameters
and we impose the same four renormalization conditions as in NLO.
We have repeated the extraction of the phase shifts and
found a negligible effect on the final result,
so that this incomplete N$^2$LO phase shift is at least as good as the one
plotted in Fig. \ref{chiraldeltaalternativefacts}.
This indicates that in the $^1S_0$ channel the effects of the N$^2$LO TPE
potential
can be compensated by a change in the strengths of our
LO and NLO short-range interactions.
Of course, this is not a full calculation of the amplitude up to N$^2$LO,
but since the change from LO to LO+NLO is small, we might expect
the iteration of NLO interactions to also produce small effects.
We intend to pursue full higher-order calculations in the future.

\bigskip
\bigskip
\bigskip
%===================================================
\section{Conclusions and Outlook}
\label{sec:conclusion}

Despite its simplicity from the computational perspective,
the two-nucleon $^1S_0$ channel has proven remarkably resistant
to a systematic expansion.
In this work we have developed a rearrangement of Chiral EFT
in this channel based on specific assumptions about
the scaling of effective-range parameters and the amplitude
zero with a single low-energy scale $M_\textrm{lo}\sim 100$ MeV.
Through the introduction of two dibaryon fields, we were
able to reproduce empirical phase shifts very well already at NLO
--- that is, including interactions of
up to relative ${\cal O}(M_\textrm{lo}/M_\textrm{hi})$ ---
from threshold to beyond the zero of the amplitude at
$k_0\simeq 340\,\mathrm{MeV}$.
The existence of a deep bound state at LO indicates that the
expansion in powers of $M_\textrm{lo}/M_\textrm{hi}$
breaks down at a scale $M_\textrm{hi}\sim 500$ MeV.

The new power counting is particularly transparent
when pions are decoupled by an artificial decrease
of their interaction strength, in which case
a version of Pionless EFT is produced.
Even in this case LO and NLO fits to empirical phase shifts look
reasonable, although the lack of pion exchange is noticeable in the
form of the energy dependence.

The apparent convergence of our LO and NLO results towards the empirical
phase shifts suggests that our PC might be the basis for
a new chiral expansion in this channel.
Our new expansion relies only on the identification of
the $N\!N$ amplitude zero as a low-energy scale.
The $^1S_0$ is unique in having such a zero and a low-energy
$S$-matrix pole --- in the $^3S_1$ channel, the amplitude zero
lies beyond the pion-production threshold, while the $^3P_0$ phase
shift crosses zero at a lower energy but displays no low-energy
pole. Moreover, both $^3S_1$ and $^3P_0$ channels are well described already 
at LO in a power counting consistent with RG invariance
\cite{NTvK,Valderrama:2009ei,Valderrama:2011mv,Long:2011qx,Long:2011xw,Song:2016ale}.

Before a claim of convergence in the $^1S_0$ channel
can be made, however, one or two higher orders
should be calculated, where additional long-range interactions appear in the
form of multi-pion exchange.
Indications already exist
\cite{Valderrama:2009ei,Long:2012ve,Long13} that two-pion exchange
and its counterterms, which enter first at N$^2$LO,
are amenable to perturbation theory in this channel.
However, it is yet to be checked whether their contributions are small enough
not to destroy the excellent agreement obtained at NLO.
This calculation is demanding because it requires treating
the NLO interaction beyond first order in distorted-wave perturbation
theory.
An incomplete N$^2$LO calculation which omits these demanding terms
suggests that higher orders might provide only very small corrections.

If this approach succeeds, then it raises new questions.
As one example, can we find an equivalent momentum-dependent
approach, which would be better suited to many-body calculations?
As another, what is the role of the quark masses in this power counting?
We have worked at physical pion mass,
but it remains to be seen how this new proposal can be implemented for
arbitrary $m_\pi$ in a renormalization-consistent manner.
We intend to address these issues in future work.

\section*{Acknowledgments}

We thank J.A. Oller and M. Pav\'on Valderrama for useful discussions.
UvK is grateful to R. Higa, G. Rupak, and A. Vaghani
for insightful comments on the role of low-energy amplitude zeros,
which have inspired this manuscript.
MSS is grateful for hospitality to the Department of Physics
at the University of Arizona,
where part of this work was carried out.
This research was supported in part by the National Natural Science Foundation
of  China (NSFC) through grant number 11375120,
by the U.S. Department of Energy, Office of Science,
Office of Nuclear Physics, under award number DE-FG02-04ER41338,
and by the European Union Research and Innovation program Horizon 2020
under grant agreement no. 654002.

%\bibliographystyle{apsrev4-1}
%\bibliography{Reference.bib}

\begin{thebibliography}{99}

%\bibitem{Beane:2010em}
%  S.R.~Beane, W.~Detmold, K.~Orginos, and M.J.~Savage,
%  %``Nuclear Physics from Lattice QCD,''
%  \textit{Prog. Part. Nucl. Phys.} {\bf 66} (2011) 1.
%  %doi:10.1016/j.ppnp.2010.08.002
%  %[arXiv:1004.2935 [hep-lat]].

%\bibitem{NPLQCD}
%S. Beane, E. Chang, W. Detmold, H.-W. Lin, T.-C. Luu, K. Orginos, A. Parre\~{n}o, M. Savage, A. Torok, and A. Walker-Loud, \textit{Phys.\ Rev.} {\bf D85} (2012) {\footnotesize 273 [arXiv:1109.2889 [hep-lat]]}.

%\bibitem{Japanese}
%T. Yamazaki, K. Ishikawa, Y. Kuramashi, and
%A. Ukawa, \textit{Phys.\ Rev.} {\bf D92} (2015) {\footnotesize 014501 [arXiv:1502.04182 [hep-lat]]}.

\bibitem{VanKolckBedaque02}
P.F.~Bedaque and U.~van~Kolck,
\textit{Ann.\ Rev.\ Nucl.\ Part.\ Sci.} {\bf 52} (2002) 339.

\bibitem{ModTheNucFor}
E.~Epelbaum, H.-W. Hammer, and U.-G. Mei{\ss}ner,
\textit{Rev.\ Mod.\ Phys.} {\bf 81} (2009) 1773.

\bibitem{Machleidt'n'Entem}
D.R. Entem and R. Machleidt,
\textit{Phys. Rept.} \textbf{503} (2011) 1.

\bibitem{Weinberg90}
S. Weinberg,
\textit{Phys.\ Lett. B} {\bf 251} (1990) 288.

\bibitem{Weinberg91}
S. Weinberg,
\textit{Nucl.\ Phys. B} {\bf 363} (1991) 3.

\bibitem{Rho:1990cf}
  M.~Rho,
  %``Exchange currents from chiral Lagrangians,''
  \textit{Phys.\ Rev.\ Lett.} {\bf 66} (1991) 1275.
  %doi:10.1103/PhysRevLett.66.1275
  %%CITATION = doi:10.1103/PhysRevLett.66.1275;%%

\bibitem{Manohar:1983md}
  A.~Manohar and H.~Georgi,
  %``Chiral Quarks and the Nonrelativistic Quark Model,''
  \textit{Nucl.\ Phys.\ B} {\bf 234} (1984) 189.
  %doi:10.1016/0550-3213(84)90231-1

\bibitem{Georgi93}
H.~Georgi,
\textit{Phys.\ Lett. B} {\bf 298} (1993) 187.
%[arXiv/9207278 [hep-ph]].

\bibitem{KSW96}
D.B.~Kaplan, M.J.~Savage, and M.B.~Wise,
\textit{Nucl.\ Phys. B} {\bf 478} (1996) 629.
%[arXiv:9605002 [nucl-th]].

\bibitem{NTvK}
A. Nogga, R.G.E. Timmermans, and U. van Kolck,
\textit{Phys.\ Rev. C} {\bf 72} (2005) 054006.
%[arXiv:0506005 [nucl-th]].

\bibitem{PavonValderrama:2005uj}
  M.~Pav\'on Valderrama and E.~Ruiz Arriola,
  %``Renormalization of NN interaction with chiral two pion exchange potential: Non-central phases,''
  \textit{Phys.\ Rev.\ C} {\bf 74} (2006) 064004.
  [Erratum: \textit{Phys.\ Rev.\ C} {\bf 75} (2007) 059905.]
  %doi:10.1103/PhysRevC.74.064004, 10.1103/PhysRevC.75.059905
  %[nucl-th/0507075.]

\bibitem{YangElsterPhillips}
C.-J. Yang, Ch. Elster, and D.R. Phillips,
\textit{Phys.\ Rev.\ C} {\bf 80} (2009) 034002.

\bibitem{Ya09B} 
C.-J. Yang, Ch. Elster, and D.R. Phillips, 
\textit{Phys.\ Rev.\ C} \textbf{80} (2009) 044002.

\bibitem{ZE12} 
Ch. Zeoli, R. Machleidt and D.R. Entem, 
\textit{Few-Body Syst.} \textbf{54} (2013) 2191. 
%[arXiv:nucl-th/0107009].
%%CITATION = EPHJA,A17,89;%%

\bibitem{Valderrama:2014vra}
  M.~Pav\'on Valderrama and D.R.~Phillips,
  %``Power Counting of Contact-Range Currents in Effective Field Theory,''
  \textit{Phys.\ Rev.\ Lett.}  {\bf 114} (2015) 082502.
  %doi:10.1103/PhysRevLett.114.082502
  %[arXiv:1407.0437 [nucl-th]].

%\bibitem{Weinberg92}
%S. Weinberg, \textit{Phys.\ Lett.} {\bf B295} (1992) {\footnotesize 114}.

%\bibitem{EntemMachleidt03}
%D. Entem and R. Machleidt, \textit{Phys. Rev.} \textbf{C68} (2003) {\footnotesize 041001 [arXiv:0304018 [nucl-th]]}.

%\bibitem{EGM05}
%E. Epelbaum, W. Gl\"ockle, and U.-G. Mei{\ss}ner \textit{Nuc. Phys.} \textbf{A747} (2005) {\footnotesize 362 [arXiv:0405048 [nucl-th]]}.

\bibitem{KSW98bis}
D.B.~Kaplan, M.J.~Savage, and M.B.~Wise,
\textit{Phys.\ Lett. B} {\bf 424} (1998) 390.
%[arXiv:9801034 [nucl-th]].

\bibitem{KSW98}
D.B.~Kaplan, M.J.~Savage, and M.B.~Wise,
\textit{Nucl.\ Phys. B} {\bf 534} (1998) 329.
%[arXiv:9802075 [nucl-th]].

\bibitem{FMS00}
S. Fleming, T. Mehen, and I.W. Stewart,
\textit{Nucl.\ Phys. A} {\bf 677} (2000) 357.
%[arXiv:9911001 [nucl-th]].

  \bibitem{Birse:2005um}
  M.C. Birse,
  %``Power counting with one-pion exchange,''
  \textit{Phys.\ Rev.\ C} {\bf 74} (2006) 014003.
  %doi:10.1103/PhysRevC.74.014003
  %[nucl-th/0507077].

\bibitem{Valderrama:2009ei}
  M. Pav\'on Valderrama,
  %``Perturbative renormalizability of chiral two pion exchange in nucleon-nucleon scattering,''
  \textit{Phys.\ Rev.\ C} {\bf 83} (2011) 024003.
  %doi:10.1103/PhysRevC.83.024003
  %[arXiv:0912.0699 [nucl-th]].

\bibitem{Valderrama:2011mv}
  M. Pav\'on Valderrama,
  %``Perturbative Renormalizability of Chiral Two Pion Exchange in Nucleon-Nucleon Scattering: P- and D-waves,''
  \textit{Phys.\ Rev.\ C} {\bf 84} (2011) 064002.
  %doi:10.1103/PhysRevC.84.064002
  %[arXiv:1108.0872 [nucl-th]].

\bibitem{Long:2011qx}
  B.~Long and C.-J.~Yang,
  %``Renormalizing chiral nuclear forces: a case study of 3P0,''
  \textit{Phys.\ Rev.\ C} {\bf 84} (2011) 057001.
  %doi:10.1103/PhysRevC.84.057001
  %[arXiv:1108.0985 [nucl-th]].

\bibitem{Long:2011xw}
  B.~Long and C.-J.~Yang,
  %``Renormalizing Chiral Nuclear Forces: Triplet Channels,''
  \textit{Phys.\ Rev.\ C} {\bf 85} (2012) 034002.
  %doi:10.1103/PhysRevC.85.034002
  %[arXiv:1111.3993 [nucl-th]].

\bibitem{Long:2012ve}
  B.~Long and C.-J.~Yang,
  %``Short-range nuclear forces in singlet channels,''
  \textit{Phys.\ Rev.\ C} {\bf 86} (2012) 024001.
  %doi:10.1103/PhysRevC.86.024001
  %[arXiv:1202.4053 [nucl-th]].

\bibitem{Song:2016ale}
Y.-H. Song, R. Lazauskas, and U. van Kolck,
%``Triton and Neutron-Deuteron Scattering up to Next-to-Leading Order in Chiral EFT,''
arXiv:1612.09090 [nucl-th].
%%CITATION = ARXIV:1612.09090;%%

\bibitem{PeripheralSinglets}
M. Pav\'on Valderrama, M. S\'anchez S\'anchez, C.-J. Yang, B. Long,
J. Carbonell, and U. van Kolck,
arXiv:1611.10175 [nucl-th].

\bibitem{Kaplan97}
D.B.~Kaplan,
\textit{Nucl.\ Phys. B} {\bf 494} (1997) 471.

\bibitem{Cohen:1998}
T.D. Cohen and J.M. Hansen,
\textit{Phys. Lett. B} \textbf{440} (1998) 233.

\bibitem{Steele:1998zc}
 J.V. Steele and R.J. Furnstahl,
  %``Removing pions from two nucleon effective field theory,''
 \textit{Nucl. Phys. A} {\bf 645} (1999) 439.
  %doi:10.1016/S0375-9474(98)00619-8
  %[nucl-th/9808022].
  %%CITATION = doi:10.1016/S0375-9474(98)00619-8;%%

\bibitem{Mehen:1999}
T. Mehen and I.W. Stewart,
\textit{Phys. Rev. C} \textbf{59} (1999) 2365.

\bibitem{Frederico:1999}
T. Frederico, V.S. Tim\'oteo, and L. Tomio,
\textit{Nucl. Phys. A} \textbf{653} (1999) 209.

\bibitem{Gegelia:1999}
J. Gegelia,
\textit{Phys. Lett. B} \textbf{463} (1999) 133.

\bibitem{Kaplan:1999qa}
  D.B. Kaplan and J.V. Steele,
  %``The Long and short of nuclear effective field theory expansions,''
  \textit{Phys. Rev. C} {\bf 60} (1999) 064002.
  %doi:10.1103/PhysRevC.60.064002
  %[nucl-th/9905027].
  %%CITATION = doi:10.1103/PhysRevC.60.064002;%%\bibitem{FS98}

\bibitem{Hyun:2000}
C.H. Hyun, D.-P. Min, and T.-S. Park,
\textit{Phys. Lett. B} \textbf{473} (2000) 6.

\bibitem{Lutz00}
M. Lutz,
\textit{Nuc. Phys. A} \textbf{677} (2000) 241.

\bibitem{Beane:2001bc}
  S.R.~Beane, P.F.~Bedaque, M.J.~Savage, and U.~van Kolck,
  %``Towards a perturbative theory of nuclear forces,''
  \textit{Nucl.\ Phys.\ A} {\bf 700} (2002) 377.
  %doi:10.1016/S0375-9474(01)01324-0
  %[nucl-th/0104030].

\bibitem{Nieves:2003}
J.M. Nieves,
\textit{Phys. Lett. B} \textbf{568} (2003) 109.

\bibitem{Oller:2003}
J.A. Oller,
\textit{Nucl. Phys. A} \textbf{725} (2003) 85.

\bibitem{ValArr:2004}
M.~Pav\'on Valderrama and E.~Ruiz Arriola,
\textit{Phys.\ Lett.\ B} {\bf 580} (2004) 149.

\bibitem{ValArr:2004bis}
M.~Pav\'on Valderrama and E.~Ruiz Arriola,
\textit{Phys.\ Rev.\ C} {\bf 70} (2004) 044006.

\bibitem{Frederico:2005}
V.S. Tim\'oteo, T. Frederico, A. Delfino, and L. Tomio,
\textit{Phys. Lett. B} \textbf{621} (2005) 109.

\bibitem{PA06}
M. Pav\'on Valderrama and E. Ruiz Arriola,
\textit{Phys.\ Rev. C} {\bf 74} (2006) 054001.

\bibitem{YangHuang}
J.-F. Yang and J.-H. Huang,
\textit{Commun. Theor. Phys.} \textbf{47} (2007) 699.

\bibitem{EntArrValMach}
D.R. Entem, E. Ruiz Arriola, M. Pav\'on Valderrama, and R. Machleidt,
\textit{Phys. Rev. C} \textbf{77} (2008) 044006.

\bibitem{SotoTarrus08}
J. Soto and J. Tarr\'us,
\textit{Phys. Rev. C} \textbf{78} (2008) 024003.

\bibitem{Shukla:2008sp}
  D. Shukla, D.R. Phillips, and E. Mortenson,
  %``Chiral potentials, perturbation theory, and the S-1(0) channel of NN scattering,''
  \textit{J. Phys. G} {\bf 35} (2008) 115009.
  %doi:10.1088/0954-3899/35/11/115009
  %[arXiv:0803.4190 [nucl-th]].


\bibitem{YangPhillips}
C.-J. Yang, Ch. Elster, and D.R. Phillips,
\textit{Phys.\ Rev. C} {\bf 77} (2008) 014002.

\bibitem{Birse:2010jr}
  M.C. Birse,
  %``Deconstructing $^{1}S_{0}$ nucleon-nucleon scattering,''
  \textit{Eur. Phys. J. A} {\bf 46} (2010) 231.
  %doi:10.1140/epja/i2010-11034-9
  %[arXiv:1007.0540 [nucl-th]].
  %%CITATION = doi:10.1140/epja/i2010-11034-9;%%

\bibitem{Harada:2011}
K. Harada, H. Kubo, and Y. Yamamoto,
%`` Pions are neither perturbative nor nonperturbative: Wilsonian renormalization group analysis of nuclear effective field theory including pions,''
\textit{Phys. Rev. C} \textbf{83} (2011) 034002.

\bibitem{AndoHyun12}
S.-I. Ando and C. H. Hyun,
\textit{Phys. Rev. C} \textbf{86} (2012) 024002.

\bibitem{Szpigel:2012}
S. Szpigel and V.S. Tim\'oteo,
\textit{J. Phys. G} \textbf{39} (2012) 105102.

\bibitem{Long13}
B. Long,
\textit{Phys. Rev. C} {\bf 88} (2013) 014002.
%[arXiv:1304.7382 [nucl-th]].

\bibitem{Harada:2013}
K. Harada, H. Kubo, T. Sakaeda, and Y. Yamamoto,
%``Convergent perturbative nuclear effective field theory,''
arXiv:1311.3063 [nucl-th].

\bibitem{Epelbaum-2015sha}
  E.~Epelbaum, A.M.~Gasparyan, J.~Gegelia, and H.~Krebs,
  %``$^{1}$S$_{0}$ nucleon-nucleon scattering in the modified Weinberg approach,''
  \textit{Eur.\ Phys.\ J.\ A} {\bf 51} (2015)
%no. 6, 
  71.
  % doi:10.1140/epja/i2015-15071-6
  % [arXiv:1501.01191 [nucl-th]].
  %%CITATION = doi:10.1140/epja/i2015-15071-6;%%
  %11 citations counted in INSPIRE as of 19 Apr 2017

\bibitem{Ren-2016jna}
  X.-L.~Ren, K.-W.~Li, L.-S.~Geng, B.-W.~Long, P.~Ring, and J.~Meng,
  %``Leading order covariant chiral nucleon-nucleon interaction,''
  arXiv:1611.08475 [nucl-th].
  %%CITATION = ARXIV:1611.08475;%%
  %2 citations counted in INSPIRE as of 19 Apr 2017

\bibitem{Long16}
B.~Long, \textit{Int. J. Mod. Phys. E} \textbf{25} (2016) 1641006.

\bibitem{Valderrama}
M.~Pav\'on~Valderrama, \textit{Int. J. Mod. Phys. E} \textbf{25} (2016) 1641007.

\bibitem{vanKolck:1997ut}
  U. van Kolck,
  %``Nucleon-nucleon interaction and isospin violation,''
  \textit{Lect. Notes Phys.}  {\bf 513} (1998) 62.
  %doi:10.1007/BFb0104898
  %[hep-ph/9711222].
  %%CITATION = doi:10.1007/BFb0104898;%%

\bibitem{VanKolck98}
U. van Kolck,
\textit{Nucl. Phys. A} {\bf 645} (1999) 273.
%[arXiv:9808007 [nucl-th]].

\bibitem{PWA}
V.G.J. Stoks, R.A.M. Klomp, M.C.M. Rentmeester, and J.J. de Swart,
\textit{Phys. Rev. C} \textbf{48} (1993) 792.

\bibitem{nnonline}
\textit{NN-OnLine}, \url{http://nn-online.org/}.

\bibitem{BeaneSavage}
S.R. Beane and M.J. Savage,
\textit{Nucl. Phys. A} \textbf{694} (2001) 511.

\bibitem{nd}
A. Vaghani, R. Higa, G. Rupak, and U. van Kolck,
in preparation.

\bibitem{Konig:2015aka}
S. K\"onig, H.W.~Grie\ss hammer, H.-W. Hammer, and U. van Kolck,
%``Effective theory of $^3$H and $^3$He,''
\textit{J. Phys. G} {\bf 43} (2016)
%no.5,
055106.
%doi:10.1088/0954-3899/43/5/055106
%[arXiv:1508.05085 [nucl-th]].
%%CITATION = doi:10.1088/0954-3899/43/5/055106;%%

\bibitem{CDD}
L. Castillejo, R.H. Dalitz, and F.J. Dyson,
\textit{Phys. Rev.} \textbf{101} (1956) 543.

\bibitem{OllerEntem}
D.R. Entem and J.A. Oller,
 arXiv:1610.01040 [nucl-th].

\bibitem{Krivoruchenko}
M.I. Krivoruchenko,
\textit{Phys. Rev. C} \textbf{82} (2010) 018201.

\bibitem{Nijm93}
V.G.J. Stoks, R.A.M. Klomp, C.P.F. Terheggen, and J.J. de Swart,
\textit{Phys. Rev. C} \textbf{49} (1994) 2950.

%\bibitem{HK82}
%H. van Haeringen and L.P. Kok, \textit{Phys. Rev. A} \textbf{26} (1982) 1218.

\bibitem{GSS08}
H.W. Grie{\ss}hammer, M.R. Schindler, and R.P. Springer,
\textit{Eur. Phys. J. A} \textbf{48} (2012) 7.

\bibitem{KoesterNistler75}
L. Koester and W. Nistler,
\textit{Z. Physik} {\bf 272} (1975) 189.

\bibitem{Lomon1974}
E. Lomon and R. Wilson,
\textit{Phys. Rev. C} \textbf{9} (1974) 1329.

\bibitem{PB09}
V.A. Babenko and N.M. Petrov,
\textit{Phys. Atom. Nuc.} \textbf{73} (2010) 1499.

\bibitem{Bedaque:1999ve}
  P.F. Bedaque, H.-W. Hammer, and U. van Kolck,
  %``Effective theory of the triton,''
  \textit{Nucl. Phys. A} {\bf 676} (2000) 357.
  %doi:10.1016/S0375-9474(00)00205-0
  %[nucl-th/9906032].
  %%CITATION = doi:10.1016/S0375-9474(00)00205-0;%%

\bibitem{Platter:2004zs}
  L. Platter, H.-W. Hammer, and U.-G. Mei{\ss}ner,
  %``On the correlation between the binding energies of the triton and the alpha-particle,''
  \textit{Phys. Lett. B} {\bf 607} (2005) 254.
  %doi:10.1016/j.physletb.2004.12.068
  %[nucl-th/0409040].
  %%CITATION = doi:10.1016/j.physletb.2004.12.068;%%

\bibitem{Stetcu:2006ey}
  I. Stetcu, B.R. Barrett, and U. van Kolck,
  %``No-core shell model in an effective-field-theory framework,''
  \textit{Phys. Lett. B} {\bf 653} (2007) 358.
  %doi:10.1016/j.physletb.2007.07.065
  %[nucl-th/0609023].
  %%CITATION = doi:10.1016/j.physletb.2007.07.065;%%

\bibitem{Contessi:2017rww}
  L.~Contessi, A.~Lovato, F.~Pederiva, A.~Roggero, J.~Kirscher,
  and U.~van Kolck,
  %``Ground-State Properties of $^{4}$He and $^{16}$O Extrapolated from Lattice QCD with Pionless EFT,''
  arXiv:1701.06516 [nucl-th].
  %%CITATION = ARXIV:1701.06516;%%

\bibitem{vanKolck:2016}
S. K{\"o}nig, H.W. Grie{\ss}hammer, H.-W. Hammer, and U. van Kolck,
arXiv:1607.04623 [nucl-th].

\bibitem{VanOers67}
W.T.H.~van Oers and J.D.~Seagrave,
\textit{Phys.\ Lett.} {\bf{24B}} (1967) 562.

\bibitem{ValderramaArriola04}
M.~Pav\'on Valderrama and E. Ruiz Arriola, arXiv:0407113 [nucl-th].

\bibitem{MA46}
S.T.~Ma,
\textit{Phys.\ Rev.} \textbf{69} (1946) 668.

\bibitem{MA47}
S.T.~Ma,
\textit{Phys.\ Rev.} \textbf{71} (1947) 195.

\bibitem{Ordonez:1992xp}
C.~Ord\'o\~nez and U.~van Kolck,
  %``Chiral lagrangians and nuclear forces,''
  \textit{Phys.\ Lett.\ B} {\bf 291} (1992) 459.
  %doi:10.1016/0370-2693(92)91404-W

\bibitem{Ordonez:1995rz}
  C.~Ord\'o\~nez, L.~Ray, and U.~van Kolck,
  %``The Two nucleon potential from chiral Lagrangians,''
  \textit{Phys.\ Rev.\ C} {\bf 53} (1996) 2086.
  %doi:10.1103/PhysRevC.53.2086
  %[hep-ph/9511380].

\bibitem{Vanasse:2013sda}
  J.~Vanasse,
  %``Fully Perturbative Calculation of $nd$ Scattering to Next-to-next-to-leading-order,''
  \textit{Phys. Rev. C} {\bf 88} (2013)
%no.4,
  044001.
  %doi:10.1103/PhysRevC.88.044001
  %[arXiv:1305.0283 [nucl-th]].

\end{thebibliography}

\end{document}